\documentclass[a4paper,fleqn,usenatbib,useAMS]{mnras}

\usepackage{threeparttable}
\usepackage{tabularx}
\usepackage{multirow}
\usepackage{graphicx}
\usepackage{svg}
\usepackage{float}
\usepackage{amsmath}	% Advanced maths commands
\usepackage{amssymb}	% Extra maths symbols
\usepackage{bm}		% Bold maths symbols, including upright Greek
\usepackage{pdflscape}	% Landscape pages
\usepackage{caption} 
\usepackage[T1]{fontenc}
\usepackage{ae,aecompl}

\usepackage{newtxtext,newtxmath}
\title{A method of photometric data extraction for asteroids from time-domain surveys}
\author[Xiaoyun Xu et al.]{
Xiaoyun Xu,$^{1,3}$
Xiaobing Wang,$^{1,2,3}$\thanks{E-mail: wangxb@ynao.ac.cn}
Karri Muinonen,$^{4}$
Antti Penttil\"{a},$^{4}$
Nanping Luo,$^{1,3}$
\newauthor
Shenghong Gu,$^{1,2,3}$
Leilei Sun,$^{1,2}$
Fukun Xu,$^{1,2}$
Yisi Liu,$^{5}$
Yue Xiang,$^{1,2}$
Dongtao Cao,$^{1,2}$
\newauthor
Jianhua Wang$^{1,3}$
\\
$^{1}$Yunnan Observatories, Chinese Academy of Sciences, Kunming 650216, China\\
$^{2}$Key Laboratory for the Structure and Evolution of Celestial Objects, Chinese Academy of Sciences, Kunming 650216, China\\
$^{3}$University of Chinese Academy of Sciences, Beijing 100049, China\\
$^{4}$Department of Physics, P.O. box 64, FI-00014 University of Helsinki, Finland\\
$^{5}$ Deep Space Exploration Laboratory, Beijing 100043, China
}

\date{}

\pubyear{2022}

\begin{document}
\label{firstpage}
\pagerange{\pageref{firstpage}--\pageref{lastpage}}
\maketitle

% Abstract of the paper
\begin{abstract}
The lightcurves of asteroids are essential for determining their physical characteristics, including shape, spin, size, and surface composition. However, most asteroids are missing some of these basic physical parameters due to lack of photometric data. Although a few telescopes or surveys are specially designed for photometric lightcurve observations of asteroids, many ground-based and space-based sky surveys for hunting new exoplanets, transient events, etc., should capture numerous small Solar System objects. This will benefit the physical studies of these objects. In order to extract data of these moving objects from time-domain photometric surveys, we have developed a new method using the model tree algorithm in the field of machine learning. A dedicated module is built to automatically identify moving objects in dataset, and extract their photometric and astrometric data.
As the first application of this novel method, we have analyzed data in five fields of the Yunnan-Hong Kong wide field photometric (YNHK) survey, from which 538 lightcurves of 211 asteroids are successfully extracted. Meanwhile, we also tested the method based on the data from NASA's Transiting Exoplanet Survey Satellite, and the result proves the reliability of our method. With derived lightcurves of 13 asteroids from the YNHK survey, we have determined their synodic spin periods, among which the periods of 4 asteroids are estimated for the first time. In future, we are going to apply this method to search for small objects in the outer part of the Solar System from the Chinese Space Station Telescope survey.
\end{abstract}


%%%%%%%%%%%%%%%%%%%%%%%%%%%%%%%%%%%%%%%%%%%%%%%%%%

%%%%%%%%%%%%%%%%% BODY OF PAPER %%%%%%%%%%%%%%%%%%

%\begingroup
%\let\clearpage\relax
%\tableofcontents
%\endgroup
%\newpage
% figure 1

\section{Introduction}
Asteroids are thought to be remnants of the planetesimals in the early stages of the Solar System: they participated in the formation process of terrestrial planets. 
The present physical characteristics of main-belt asteroids (MBAs) can help us to understand how terrestrial planets formed and give hints about the origin and the evolution history of themselves. Near-Earth objects (NEOs) mainly originate from the asteroid main belt and pose an impact hazard to the Earth, with potential for damaging the environment and even threatening the human civilization.

Just like Michel et al. mentioned \citep{AsteroidsIV}, a dramatic increase in the number of discovered asteroids has occurred in the last thirty years, due to the application of charge-coupled device (CCD) technology in asteroid surveys, e.g., LONEOS, LINEAR, NEAT, NEOWISE, Spacewatch, Pan-STARRS, and Catalina Sky Survey \citep{Spacewatch,LONEOS,CSS,NEAT,LINEAR,Pan-STARRS,NEOWISE}. At present, more than one million asteroids have been discovered. However, the number of asteroids with known physical characteristics (size, shape, spin parameters, surface photometric parameters, etc.) is much smaller. The best photometric data collection of asteroids, Asteroid Lightcurve Data Base \citep[LCDB,][]{WARNER2009134}, had 373,819 lightcurves of 23,989 asteroids in June 2022. So far, the spin periods of more than 8,300 asteroids are determined and about three thousand asteroids have shape parameters and orientation information \citep{2010A/&A...513A..46D}. The SDSS provided multi-bands photometry for plenty of asteroids \citep{Ivezi2001}, with those data the composition type of 400 thousands asteroids were analysed \citep{Sergeyev2021}. Also, \cite{Sergeyev2022} extracted multi-filter asteroid photometry from the SkyMapper survey, and more data could be used in the taxonomic classification of asteroids. More than five hundred asteroids were measured to derive their phase function utilizing the data from the Gaia survey \citep{Martikainen2021,Muinonen2022}. The main reason for the lack of the physical parameters of asteroids is that sufficient photometric data covering a range of rotational phases and aspect angles are required, and this is laborious to realize. Normally, the observations for obtaining such photometric data need to span a long period of time.

Nowadays, there are many sky surveys aimed at searching for new celestial objects and phenomena, such as exoplanets and transient events, e.g., CHEOPS, WASP, HATNet, Kepler, TESS \citep{HATNet,WASP,Kepler,CHEOPS,TESS}. These surveys stare at the target sky areas for a certain period, e.g., whole night or even longer period by multiple exposures, thus are called time-domain surveys. A time-domain survey could provide time-series photometric data of celestial objects in the targeted sky regions, and serendipitously capture Solar System bodies. As a result, time-domain surveys are excellent sources for asteroid lightcurves. However, for a time-domain survey not dedicated to asteroids, the data reduction pipeline usually does not include functions to identify and extract data of small Solar System objects (SSOs).

The development of computer technology significantly improves the efficiency of searching and identifying asteroids in sky surveys. There have been several different methods to exploit moving objects in sky surveys. It is efficient to search for known asteroids by relying on their updated ephemeris data  corresponding to observational epochs. This method was successfully used in the Gaia mission \citep{spoto2018gaia}, VISTA-VHS survey \citep{Popescu2016} and WTS survey \citep{10.1093/mnras/stz2727}. Another method, called the blind method, can detect and identify moving objects in certain sky area without prior information. Spacewatch employed the moving object detection program \citep{Spacewatch} and the near-Earth asteroid tracking (NEAT) program used this method to search for new moving objects by fitting a linear motion of at least three detections in one night \citep{NEAT}. The Pan-STARRS, NEOWISE, LSST, and ATLAS surveys, with the idea of blind search, used the variable kd-tree algorithm to link detections within and between nights \citep{NEOWISE,Pan-STARRS-System,LSST,ATLAS}.

We have been running a time-domain survey, Yunnan-Hong Kong wide field photometric (YNHK) survey, since 2016 \citep{gu2022}.  For the YNHK survey, besides its main goal to discover new transiting exoplanets, other research topics are also under consideration, including variable stars \citep{10.1093/mnras/stac1406,2022AJ....163..167L} and asteroids. We have already established a pipeline for the data reduction of stellar objects. Here, we introduce a module for the abovementioned pipeline to mine the asteroid data from the YNHK survey. Recently, machine learning techniques are widely used in astronomy, e.g., in data mining and classification. \cite{Kruk2022} used deep learning algorithms to explore asteroid trajectories from Hubble Space Telescope observational images. \cite{Erasmus2019} applied the PHOTOMETRYPIPELINE \citep[PP,][]{Mommert2017A&C} to extract the photometric data from the Korea Microlensing Telescope Network. In order to exploit asteroids' lightcurves efficiently and automatically from the YNHK survey, we have developed a machine learning based method to identify asteroids in the targeted fields, and to extract their lightcurves and astrometric positions, plugged in the main body of the pipeline. In our method, several algorithms are applied, such as model regression and decision tree.

Another motivation to develop our method is to mine small SSO data from the planned sky survey of the Chinese Space Station Telescope (CSST). The CSST, a 2.0-m space telescope, will be launched in 2024 to orbit with the Chinese Space Station, and aims to be operated for at least ten years \citep{zhan2021}. The CSST will be equipped with five instruments: survey module, terahertz receiver, Multi-Channel Imager (MCI), Integral Field Spectrograph (IFS), and Cool Planet Imaging Coronagraph (CPI-C). During its operation, the survey module will occupy 70 \% of whole observational time. The CSST survey observations of ten years will cover 17,500 deg$^2$ sky area with the limiting magnitude brighter than 26~mag in the r band, and 400 deg$^2$ deep-sky areas with a limiting r magnitude brighter than 27.2~mag. Considering the capabilities of the CSST, more small SSOs will be observed during the survey.

The structure of the paper is arranged as follows: we introduce the observation strategy and pipeline for data reduction of the YNHK survey in Sec.~2. The description of our novel method for moving object detection is presented in Sec.~3. In Sec.~4, the application results in the YNHK survey are shown. The limitations and extensions of our method are discussed in Sec.~5. Finally, a brief summary is given in Sec.~6.

\section{YNHK survey and data reduction}
\subsection{YNHK survey}
The YNHK survey is an ongoing project under collaboration between Yunnan Observatories, Chinese Academy of Science (CAS) and Hong Kong Astronomical Society (HKAS), China. The YNHK survey uses a Centurion 18-inch telescope located at the Lijiang station of Yunnan observatories (observatory code O45). This telescope is equipped with an Apogee ALTA U16 camera with a 4K $\times$ 4K CCD chip, corresponding to a FoV of 1.67$^{\circ}$ $\times$ 1.67$^{\circ}$ with a resolution of 1.47"/pixel. The R filter and clear filter are set up in this telescope, and for the YNHK survey, the clear filter was used for the observations. On average, the seeing we estimated by the FWHM (Full width at half Maximum) of stellar images of YNHK survey is 1.2". \cite{Xin2020} estimated an averaged seeing of 1.0" for the Lijiang station.

The YNHK survey is dedicated to discovering new transiting exoplanetary systems, its strategy is to monitor some selected sky regions for time spans as long as possible, and is also used to find new variable stars.
Six sky areas distributed at different ecliptic longitudes are chosen and named as Jan-Feb, Mar-Apr, May-Jun, Jul-Aug, Sep-Oct, and Nov-Dec. Each sky area is composed of 4 fields and is observed continuously for 3 months every year. 2--3 sky areas are observed with a fixed exposure time of 8 seconds each night, which allows the photometric data to have short cadences. For more details, see \cite{gu2022}.

In practice, the distribution of observations in each year is not uniform, the most frequent observations are obtained in the winter and spring seasons, while the observations in the summer season are often interrupted due to rainy weather. The limiting magnitude can reach 17.5~mag in the V band with a signal-to-noise ratio of 5, thus at least ten thousand celestial objects, thirty thousand for some fields, can be detected in one image.

According to typical motions of MBAs and NEOs (32"/hour and 43"/hour), the visible times of asteroids in our FoV can reach 7 and 5 days, respectively. That means we can obtain multiple lightcurves for a serendipitously observed asteroid. Such photometric data will be helpful for physical studies of asteroids, at least the spin periods can be determined for targets missing spin parameters.

The YNHK survey is not dedicated to asteroids or other small SSOs, so the first version of data reduction pipeline is only suitable for the stellar sources in the observed images. Based on the main body of our data reduction pipeline, we add a module with the aid of machine learning techniques to automatically fulfil the moving object identification and data extraction. 

%2.2
\subsection{Data reduction pipeline}
\begin{figure}
	\includegraphics[width=\columnwidth]{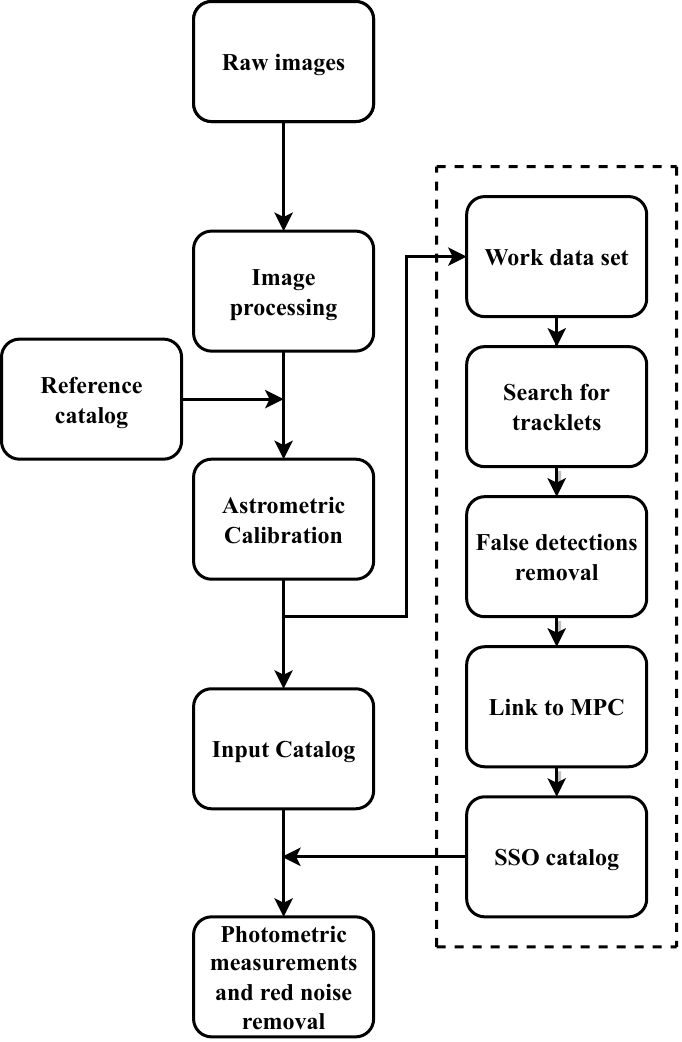}
    \caption{The pipeline for the YNHK survey. The part within the dashed lines is used to search for moving objects.}
    \label{YNHKpipe}
\end{figure}

For the stellar objects, the first version of the data reduction pipeline of the YNHK survey contains four parts: raw image processing, astrometry calibration and cross-match with the catalog, photometric measurement and red noise correction, and search for special features in time-series photometric data. This pipeline is developed by us using Python language, IRAF, and pyraf interface \citep{gu2022}.  The work flow of the pipeline is shown in Fig. \ref{YNHKpipe}.

During data processing with the first version of the pipeline, the selected stellar sources by the cross-match with certain reference catalog in the images are extracted automatically according to the scientific goal. As for moving objects in the images, we insert a module between the steps of astrometry calibration and photometry measurement. In this section, we will simply introduce key points of the first version of the pipeline and the idea to insert an additional module related to moving object detection. 

\begin{figure}
	\includegraphics[width=\columnwidth]{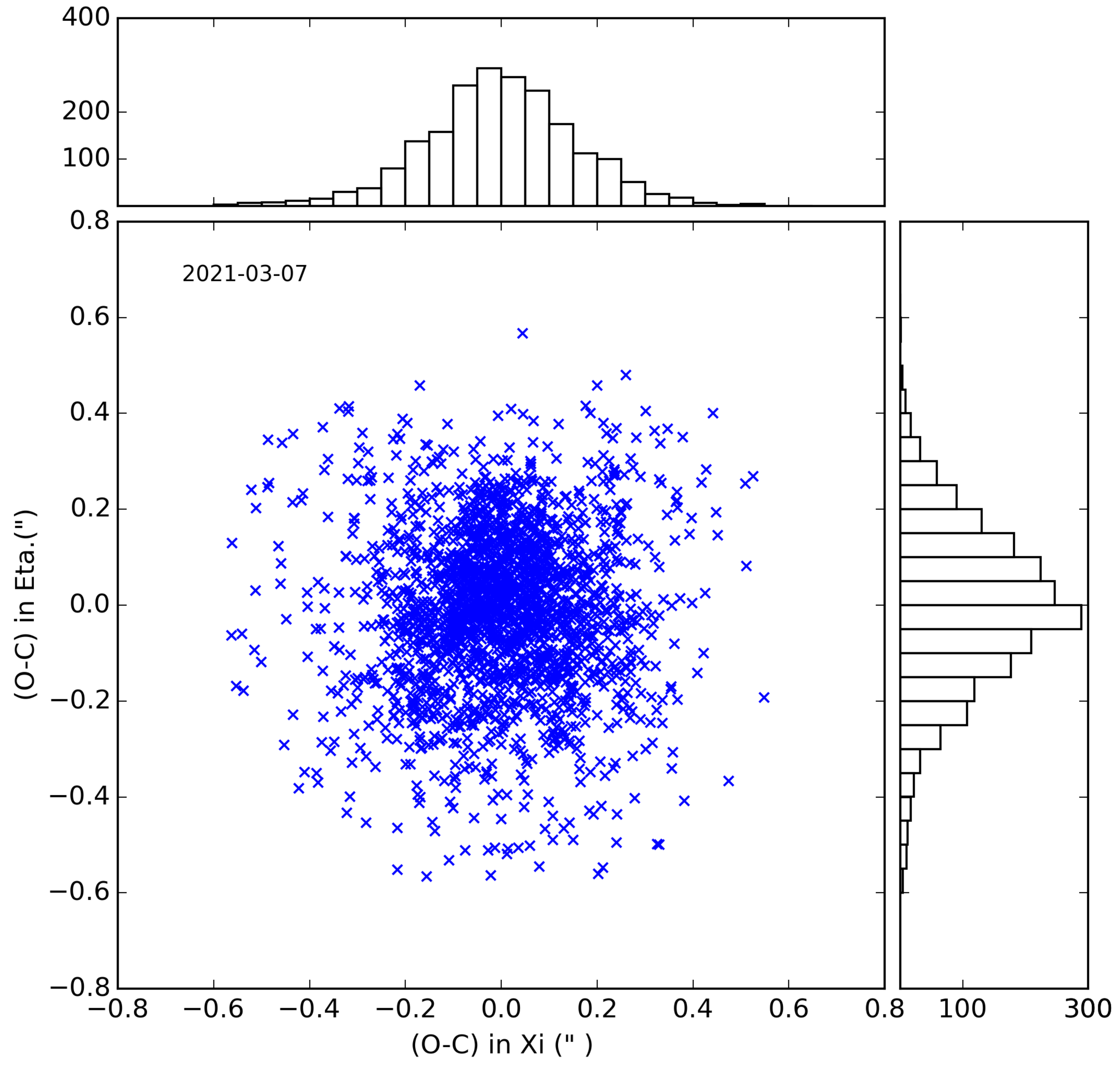}
    \caption{The distribution of astrometric residuals for the data observed on 2021.03.07.}
    \label{cal_pos}
\end{figure}

To fulfil automatic target identification of the YNHK survey, we carry out an astrometric calibration for a selected image with the best seeing condition in a given night with the help of the Gaia EDR3 catalog \citep{GaiaEDR32021}. The procedure of the astrometric calibration is to establish a relationship between the physical pixel positions in the CCD image and their astrometric coordinates in a given catalog. Here, we apply two 13-parameter functions (see eq. \ref{eq:quadratic}) to express the relationships between the Gnomonic projection quantities of the right ascension (R.A.) and declination (Dec.) of sources on the sky plane ($xi$, $eta$) and the CCD pixel positions ($x$, $y$):
\begin{equation}
    \begin{split}
        \begin{aligned}
        &xi(x,y)=&(a_1,\ldots,a_{13}) \ (x,y,\sqrt{x^2+y^2},x y,x^2,y^2,\\
        && x^2y,x y^2,x^3,y^3,x^3y,x y^3,x^2y^2)^T,\\
       &eta(x,y)=&(b_1,\ldots,b_{13}) \ (x,y,\sqrt{x^2+y^2},x y,x^2,y^2,\\
        && x^2y,x y^2,x^3,y^3,x^3y,x y^3,x^2y^2)^T.\\   
        \end{aligned}
    \end{split}
	\label{eq:quadratic}
\end{equation}
In practice, the procedure of astrometric calibration is an iterative process. In the beginning, we use the webtool of \emph{Astrometry.net} \citep{2010AJ....139.1782L} to get initial input data corresponding to the image, which contains the pixel positions and astrometry coordinates of the matched stellar sources in the uploaded image. Then, we perform the least-squares fit for the initial data with the afore described polynomial function of 13 parameters, called as plate parameters of the CCD. Using derived relationships, all celestial objects detected in the image are cross-matched to Gaia EDR3 catalog by a selected threshold, then astrometric position of those matched stars from EDR3 are used to update the parameters of CCD. Above iteration procedure is running until the number of matched sources in the image is not changed significantly. The threshold to cross-match these detected objects to Gaia catalog is set as the 3 times of RMS which is the square root of sum of the standard deviations of (O-C) in $xi$ and $eta$. On average, the involved reference stars (matched objects brighter than 17.0 mag in the V-band) in observed images of the YNHK survey are more than 2,000. As an example, Fig. \ref{cal_pos} shows (O-C) errors (differences between the observed positions of stars and that in Gaia EDR3) of astrometry calibration on March 07, 2021. The means of (O-C) in $xi$ and $eta$ are -0".002 and -0".009 with standard deviations of  0".16 and 0".17, respectively.

Once the astrometric solution of one observing night is derived, it is easy to do the mutual conversion between the pixel positions of celestial objects in an image and coordinates in the J2000.0 ICRF frame. Setting an input catalog for each observed field, we can automatically carry out the photometric measurements for selected objects listed in the input catalog. The input catalog for the identification of asteroids is generated from Gaia EDR3, the magnitude limit of selected stellar sources reaches 18 for the case of better weather conditions.

Based on the input catalog of an observed field, the pixel positions of selected celestial objects in each image are calculated by their astrometry coordinates derived by the astrometric solution. The instrumental magnitudes of selected objects are measured using the APPHOT task of the IRAF package. In order to get precise photometric data of the targets, the optimal aperture is chosen by comparing the scatter of lightcurves. The time stamp of observation is corrected to the mid-time of the exposure. In order to enhance the signal-to-noise ratios of light variations, the coarse de-correlation method \citep{collier2006fast} and SYSREM method \citep{tamuz2005correcting} are applied to correct systematic errors in the photometric data. During the iterative analysis of the de-correlation procedure, the variance $\sigma_t^2$ on each image caused by transient phenomena and the intrinsic variance $\sigma_s^2$ of each source are estimated. The reference stars are selected by a threshold on values of $\sigma_s$. During the procedure of the SYSREM method, the systemic effects caused by air mass and moonlight are simulated with the reference stars and removed for all targets in the input catalog.

\subsection{Asteroids in the YNHK survey}
\begin{figure}
	\includegraphics[width=\columnwidth]{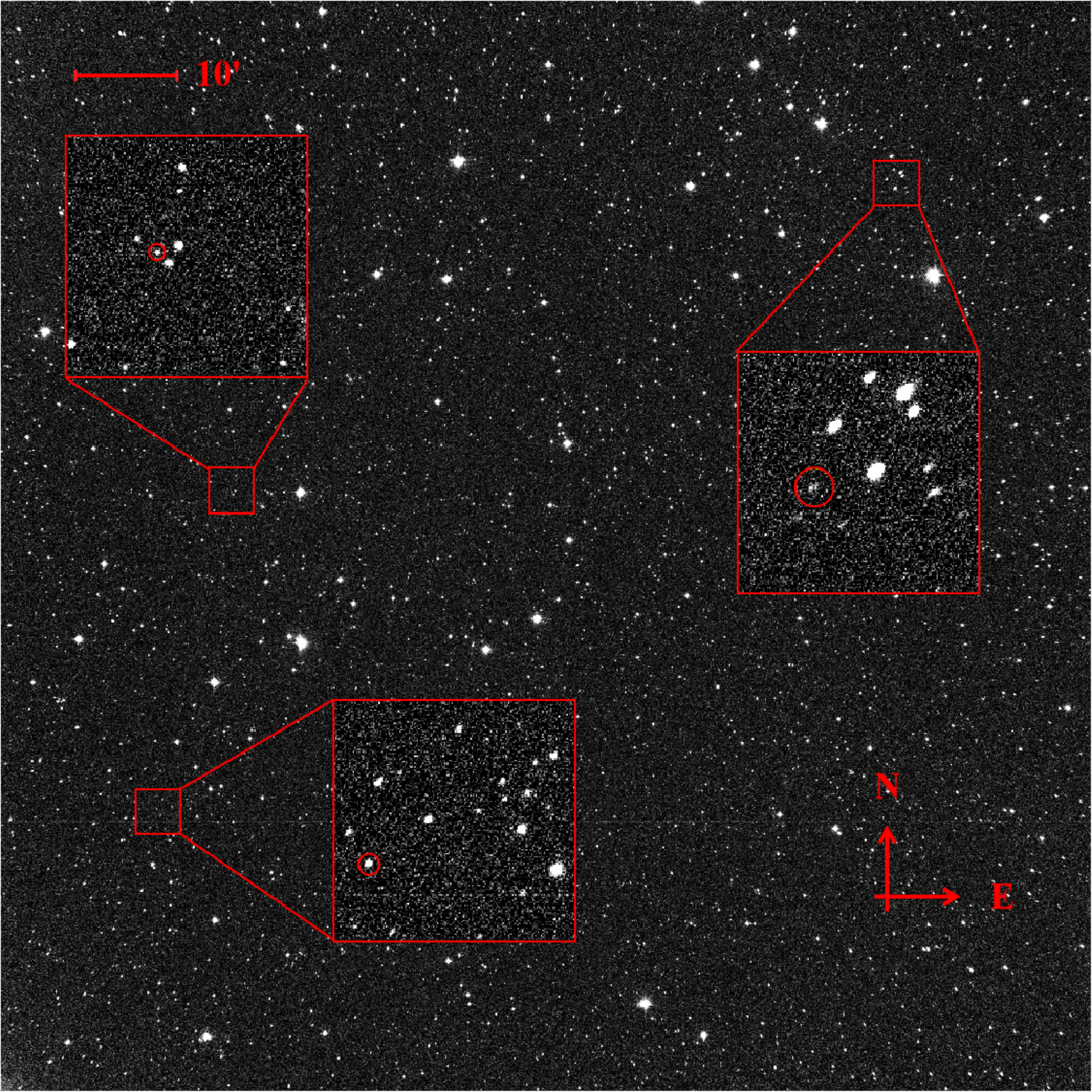}
    \caption{An example of a CCD image in sky area of May-Jun, in which asteroids (18862) Warot, (13739) Nancyworden, and (8007) 1988 RU6 are marked by red circles.}
    \label{Schematic}
\end{figure}

Although the YNHK survey is not dedicated to observing asteroids, its merits, i.e., wider FoV, short exposure time, and long observation period, will provide us with photometric data of thousands of asteroids. The typical photometric data of asteroids passed in the YNHK survey for one night corresponds to a lightcurve spanning about 4 hours, depending on the observing strategy of the survey. Sometimes, photometric data in one night can span for less than 4 hours due to weather conditions. Such photometric data of asteroids can still be beneficial for photometric studies, at least for determining the synodic periods of asteroids. For the imaging observations of 8-s exposure, asteroids show features of point sources in the CCD images, an example with asteroids marked by red circles is shown in Fig. \ref{Schematic}. Considering the point source features of asteroids and the efficiency of data reduction of moving objects, we have developed a novel method based on the machine learning technique to conduct a blind search for asteroids passing in our observed fields and to obtain their lightcurves and astrometric data.

\section{Data processing of asteroids}

Generally, data reduction of small SSOs in a sky survey should contain the identification of moving objects in the observed images and linking them to known SSOs. For long exposure observations in a stellar tracking mode, moving objects can be discerned by elongated images. For short-exposure observations, significant positional shifts relative to neighboring stars in at least 3 observations are used to identify them in one night.

For the case of the YNHK survey, the data processing of asteroids is based on the 'text images' of a field in one observational night. A text image lists the information of  sources (i.e. observation time, pixel position, flux, et al.) on a CCD image detected by the Daofind task of the IRAF software assuming a threshold of flux ratio of signal to noise. The data processing is composed of four steps (see Fig. \ref{YNHKpipe}): prepare the work dataset, search for tracklets of moving objects, remove false detections from the tracklets, and link to known asteroids. The steps of automatic searching for the tracklets and removing the false detections are carried out by a revised decision tree and isolation forest algorithms, respectively. The module is built using Python language.

\subsection{Prepare work dataset}\label{WD}
The YNHK survey is a time-domain photometric survey, and provides us with long time-series imaging data for selected sky areas. 
For the 4K $\times$ 4K images of the survey, the differential image method and stacking images method to identify the asteroids would be time-consuming. Furthermore, it is not easy to integrate the above-mentioned methods into the pipeline in the YNHK survey. For economical and efficient processing of asteroid data, we developed a module that is easy to be inserted into the pipeline of the survey. This module works on the text images of an observed field in one night or longer period. 

The text images of an observed CCD image are generated from the output file of the DAOFIND task of the IRAF package. For searching for the asteroids passing in the YNHK survey fields, a threshold of 5 sigma of the sky background is set when running the DAOFIND task. On average, sources brighter than 17.5 mag in the V band can be detected in an image, and the number of detected sources reaches tens of thousands. Thus, a text image file, in which one line corresponds to one detected source called \emph{detection}, involves tens of thousands of detections. For each detection, some information, such as the time stamp of observation $t$ and pixel position of the detected source ($x$,$y$), are selected to make the text image. Additionally, the R.A. and Dec. coordinates ($\alpha$,$\delta$) (in the frame J2000.0 ICRF) of all detections are calculated by the derived astrometric solution in the observing night, and added into the text image. At present, five features, the components ($t$, $\alpha$, $\delta$, $x$,$y$), are involved for each detection.

\begin{figure}
	\includegraphics[width=\columnwidth]{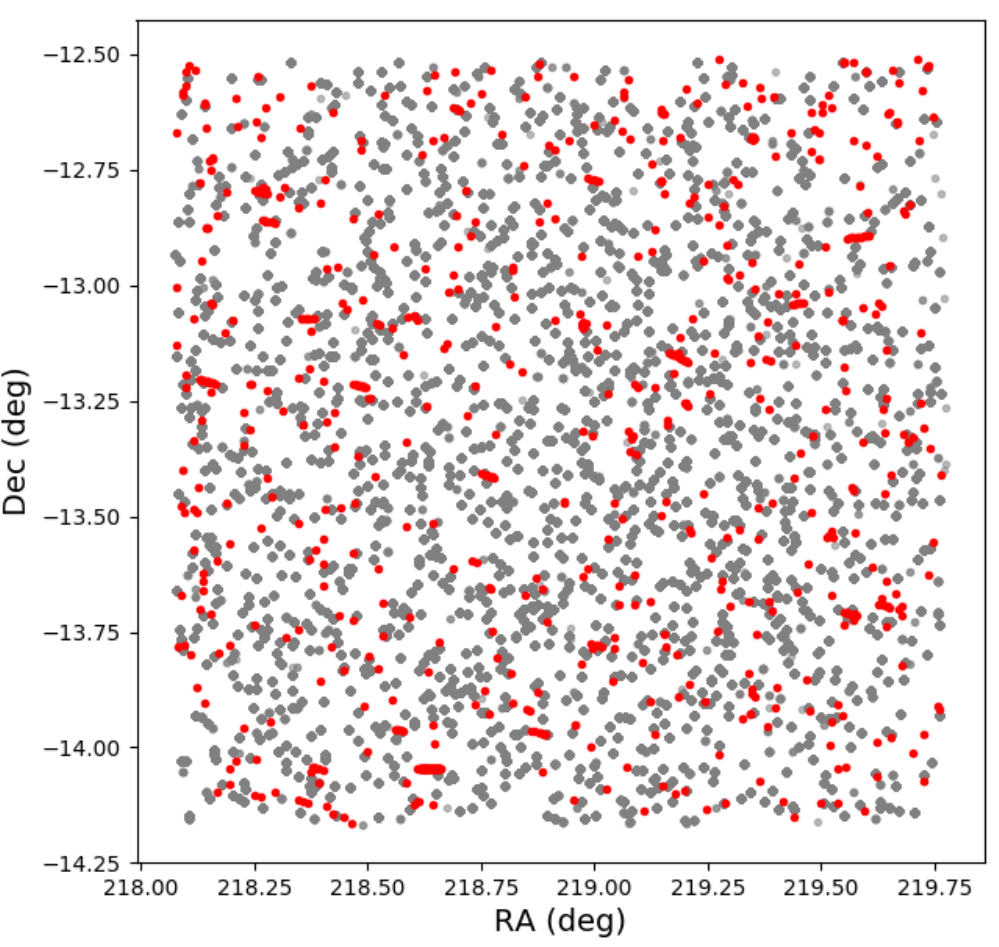}
    \caption{The distribution of astrometric positions of detections on 2019 May 10. The grey dots and red dots are detections from the raw dataset and the work dataset, respectively. }
    \label{dataset}
\end{figure}

The work dataset used to identify the passed asteroids is formed by stacking all text images of the same field in a specific period, i.e., one night, or longer time. The 'stacked' text images, called as the raw dataset, may contain millions of detections if 100 images are involved. For example, Fig. \ref{dataset} shows the ($\alpha$,$\delta$) map of detections in the raw dataset with grey dots.

For speeding up the identification of moving objects, the raw dataset is thinned by removing the stellar sources. The idea to remove the stellar sources from the raw dataset is simple. The sources detected in multiple images, but with very close astrometric positions are first removed. Considering the RMS of the astrometric calibration of the YNHK survey is around 0".15 in both R.A. and Dec., we pick and remove stellar sources by a position shift threshold of 0".2.  Again, we use the Gaia EDR3 catalog to identify and reject the rest of the stellar objects in our raw dataset by comparing the astrometric coordinates of each detection to that in this catalog with a threshold of 0".2.

After dealing with the raw dataset, the number of detections remaining is less than 5 \%. The reduced raw dataset is called the work dataset. The work dataset, shown as the red dots in Fig. \ref{dataset}, contains faint objects, residual cosmic rays, bad pixels, some other artificial signals, and moving objects occasionally passing the field of the survey. The identification of asteroids is carried out in the work dataset.

\begin{figure}
    \centering\hfill
	\includegraphics[width=\columnwidth]{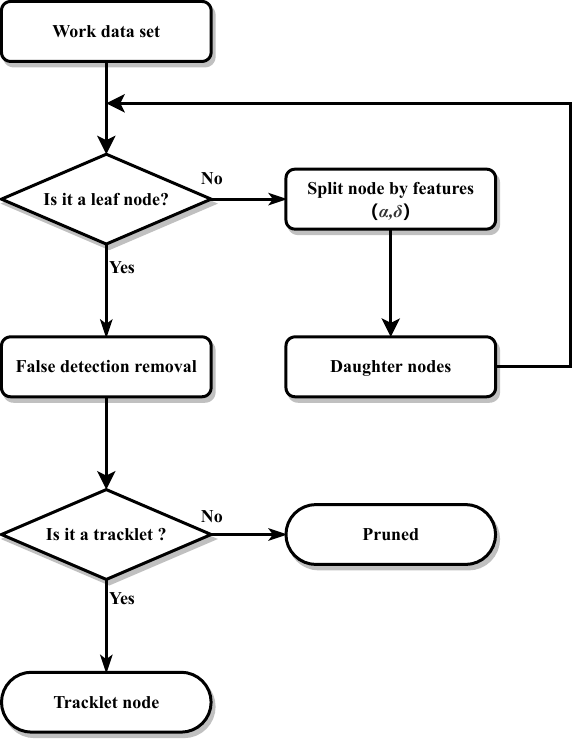}
    \caption{The work flow for searching tracklets. }
    \label{MT}
\end{figure}

\subsection{Search for tracklets of asteroids}\label{TIM}
According to the definition in \cite{2005PhDT........14K}, the collection of data points which constitutes the motion trajectory of a moving object, is called a \emph{tracklet}. The key step for extracting the asteroid data from the YNHK survey is to automatically identify the tracklets of moving objects. For this goal, we apply a machine learning technique to run a blind search for the tracklets of moving objects in the work dataset.  According to the observing strategy of the YNHK survey, we can derive continuous observations of about 4 hours for the selected fields each night.  Considering the range of motion rates of MBAs and NEOs ( 30"--40" per hour, 25"--80" per hour, respectively), a low-order approximation is adopted to represent the trajectories of asteroids in the YNHK survey for one night. Here, a linear function of time $t$ is used for MBAs and NEOs, and a quadratic function of time for the cases of fast moving NEOs or space debris. The task to identify the moving objects is to search tracklets in the work dataset, which are represented with models of their motions in R.A. and Dec. directions (see Eq.~(\ref{eq:models})):
\begin{equation}
    \begin{split}
        \begin{aligned}
            &\hat{\alpha}=\alpha_0+w_1t+w_2t^2, \\ 
            &\hat{\delta}=\delta_0+w_3t+w_4t^2. \\ 
        \end{aligned}
    \end{split}
	\label{eq:models}
\end{equation}

To search for tracklets in the work dataset, we apply the model tree algorithm and a proposed pruning method. The former is one of the advanced versions of the decision tree technique \citep{10.5555/2361796} of machine learning, and the latter is proposed for identifying the tracklets in the leaf nodes of the derived model tree.  

\begin{figure*} 
\xdef\xfigwd{\linewidth}
    \centering
      \includegraphics[width=\linewidth]{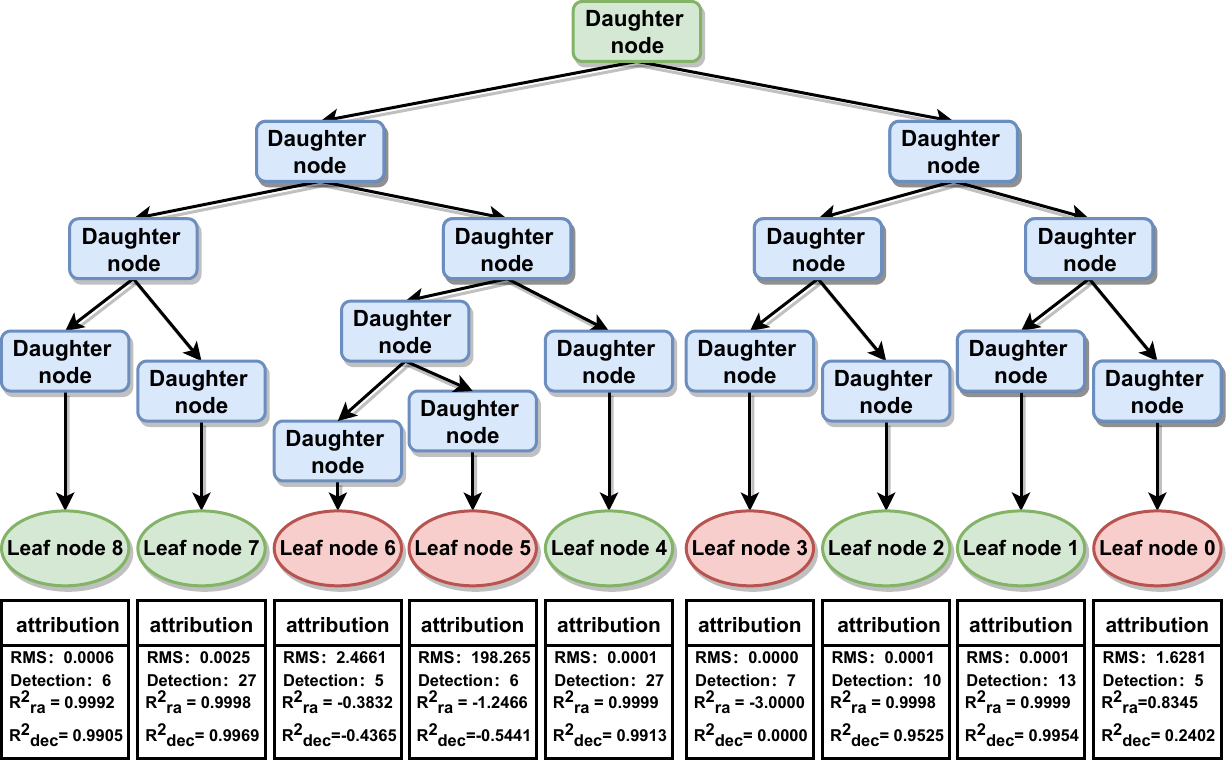}
	  \caption{One part of the model tree structure for the data  observed on 2019.05.10.}
	  \label{tree} 
\end{figure*}

The decision tree algorithm is widely applied in classification, regression, and clustering methods. By the decision tree algorithm, the input dataset is split recursively into smaller segments until all of the samples are of the same class or following a model. Such popular algorithms include the Classification And Regression Trees \citep[CART,][]{breiman1984classification}, Iterative Dichotomiser 3 \citep[ID3,][]{quinlan1986}, and C4.5 algorithm \citep{Quinlan1993}. The main difference among the three algorithms is the criterion used to determine the pattern of split to a dataset. The 'entropy' of information is frequently used as a criterion for the classification problem. Concretely, CART, ID3, and C4.5 use the gini index, information gain, and gain-ratio as their criteria, respectively.

The model tree algorithm is an advanced version of the CART algorithm, which recursively splits the dataset into small segments output for regression models. For our case, the recursive splits to the work dataset only happen on features $\alpha$ and $\delta$, that means the observed sky field is recursively split into small segments or leaf nodes, and possible tracklets will be found in the leaf nodes. For each binary split to a dataset, there is a cost function used as the criterion to select the best feature and the threshold value. Here, we apply the cost function $J(K,V_K)$ calculated with Eq.~(\ref{eq:costfunction}), which is the sum of residual sum of squares (RSS) of the two daughter nodes:

\begin{equation}
    \begin{split}
        \begin{aligned}
            &J(K,V_K)=m_{\mathrm{left}} \emph{MSE}_{\mathrm{left}}+m_{\mathrm{right}} \emph{MSE}_{\mathrm{right}} \ \ \textrm{and}\\
            &\emph{MSE}_{\mathrm{node}}=\frac{1}{m}\sum_{i=1}^{m} (y_{i}-\hat{y}_{i})^{2} \ ,\\
        \end{aligned}
    \end{split}
	\label{eq:costfunction}
\end{equation}
where $K$ is the feature index in our work dataset, $K=1$ stands for R.A. and 2 for Dec. $m$ and $\emph{MSE}$ are the number of samples in a node and the mean squared error. Accordingly, $y_{i}$ and $\hat{y}_{i}$ are the input data points in a node and the corresponding predicted values given by the regressed model. 

The construction of the model tree starts from the root node, which is the work dataset in our method. The procedure of recursive split to the work dataset will stop when it meets a specific condition. Following the algorithm, the stopping condition we used includes two quantities: TolN, the minimum number of data points included in a split, and TolS, the tolerance on the error reduction for a split. If the number of data points in a node is less than TolN, this node will never be split, and taken as a leaf node. 
%For a node, no best feature and the threshold is found to reduce the cost function at least by the value of TolS, which is also taken as a leaf node.
When binary splitting a node, no best feature and threshold are found to decrease the cost function of that node at larger than TolS, thus this node is taken as a leaf node. The schematic diagram of the procedure of tracklets search with the model tree algorithm is shown in Fig. \ref{MT}. For the YNHK survey, we consider the RMS of the astrometric calibration and source extraction process, 0.01$^{\circ}$,  as a proper value for TolS. TolN is set to 6 to ensure that enough data points are involved in the regression of the linear model.

For the derived model tree, we check the leaf nodes according to the characteristics of tracklets of asteroids. Generally, we derive two kinds of leaf nodes with the model tree algorithm, one is derived by the criterion of TolN, and one by the criterion of TolS. For the leaf nodes, we check the RMS and the correlation coefficient $R^2$ of the regression model in the leaf. Only the node with RMS less than 0.01 and both $R^2$'s in R.A. and Dec. larger than 0.9 will be kept. Otherwise, the node will be pruned. Therefore, the final model tree only contains leaf nodes with  possible tracklets of the small SSOs.

As an example, we show the structure of one part of the model tree for a single field's observational data on 2019 May 10 in Fig. \ref{tree}. The green rectangle is the root node, blue rectangles represent daughter nodes, and the ovals are the leaf nodes. The leaf nodes with green color bear tracklets of small SSOs, and those of red color do not include tracklets of small SSOs.

\begin{figure} 
\xdef\xfigwd{\textwidth}
    \centering
      \includegraphics[width=\columnwidth]{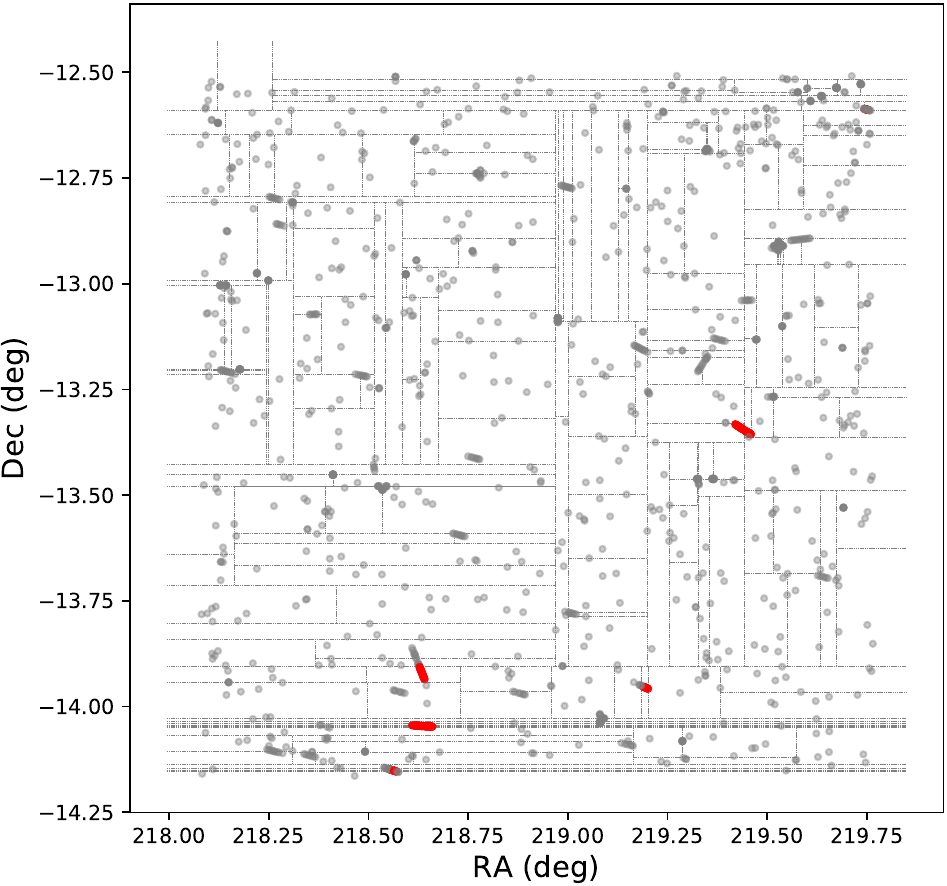}
	  \caption{The partitions in ($\alpha$,$\delta$) space created by the model tree procedure for the work dataset on 2019.05.10. The real tracklets in leaf nodes are marked by red dots.}
	  \label{2Dtree} 
\end{figure}

From Fig. \ref{2Dtree} we note that some leaf nodes may not be real tracklets of the small SSOs, or the tracklet of an SSO is separately located in additional leaf nodes, and some have no tracklets. To find the real tracklets of the SSOs, we filter again the remaining leaf nodes by the values of the linear coefficients of the regression models referencing the proper motion rates of the small SSOs in R.A. and Dec. directions.

\begin{figure} 
\xdef\xfigwd{\columnwidth}
    \centering\hfill
      \includegraphics[width=\columnwidth]{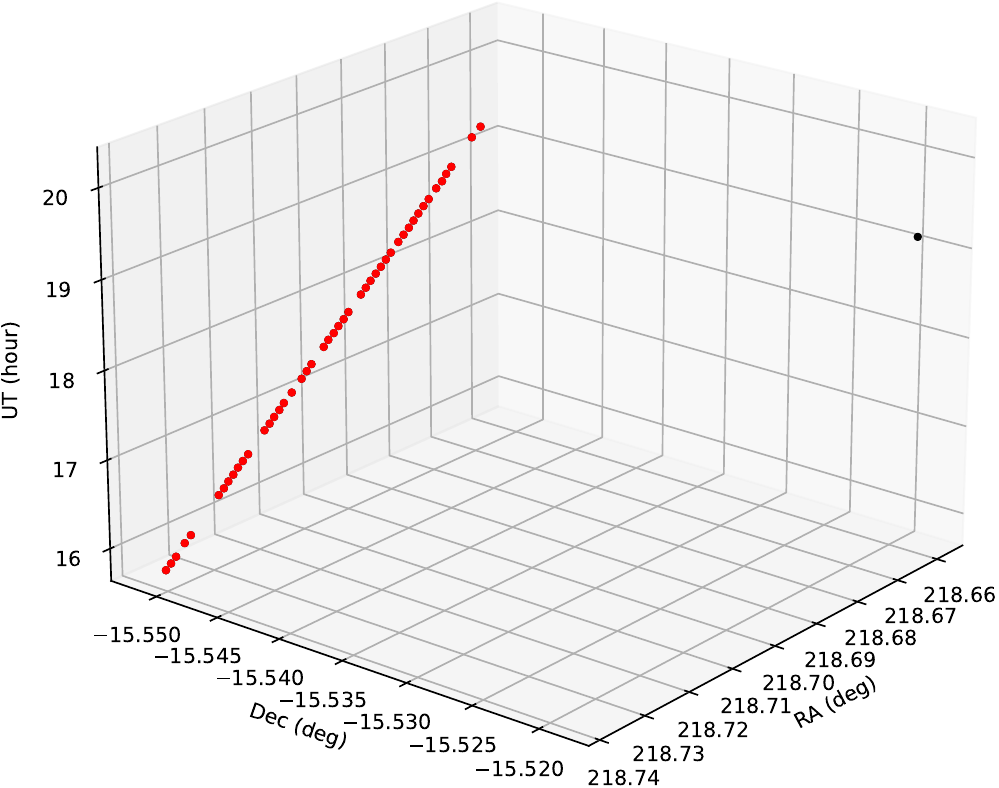}
	  \caption{An example of a leaf node with a tracklet. The red dots are detections of an asteroid, the black dot is the noise data point.}
	  \label{if} 
\end{figure}

Furthermore, we find that the leaf nodes with real tracklets also contain some data points which do not belong to the SSOs (black dots in Fig. \ref{if}). We take those data points as noise and they can decrease the precision of model regression. For removing the noise in the leaf nodes of tracklets, we apply the isolation forest algorithm.

The isolation forest is an unsupervised machine-learning method and widely used in anomaly detection. An isolation tree (ITree) is derived by recursive and random binary split to a dataset until an individual data point belongs to a leaf node, or the maximum depth of the tree reaches a given threshold. An anomalous data point quickly reaches a leaf node in an ITree, so a shorter branch is indicative of anomaly. Here, a number of isolated trees (i.e., $n_t$) are generated upon the dataset, then the ''anomaly score'' for every given point is calculated by its averaged path length over $n_t$ ITrees and a normalizing factor. For a more detailed description of the mentioned criteria, one can refer to \cite{4781136}. 

In our case, we individually analyze each leaf node by the isolation forest algorithm and remove 10 \% of data points with the lowest anomaly score in each node. The model regression for the leaf nodes of tracklets is re-calculated by robust regression. 
Finally, the leaf nodes of the model tree are updated.

\subsection{Link the tracklets to the Minor Planet Center}
After the above steps, the derived model tree can tell us whether moving objects of the Solar System pass our observed fields. By the number of remaining leaf nodes and the parameters of model regression, we can know how many and where they are in the observed images. Then, the astrometric positions and proper motions of the detected small SSOs are uploaded to the Minor Planet Checker service of the Minor Planet Center (MPC). If the uploaded positions of the detected moving object are close to the predicted ones of a known object by MPC in the range of 60", and modeled proper motion also match within certain limits, the detected source by the YNHK survey is associated with the MPC successfully. Tracklets with no corresponding objects found by the linkage procedure are probably new small SSOs and will be marked with observational date, field number, and tracklet index.

Sometimes the trajectory of a moving object in one night is separated into multiple segments, mainly because of brightness dimming of moving object at some moment for, i.e., weather conditions, moonlight, or even its intrinsic variation. Analyzing the velocity and link results of the tracklets, the tracklets belonging to the same asteroid will be combined into a single tracklet.

For all detections of identified small SSOs, the pixel positions at each images are re-calculated according to their astrometric position at the corresponding time stamps predicted by the regression models in the leaf nodes. After that, the small SSO catalog contained pixel and astrometric coordinates of the detected moving objects is added into the photometric input catalog. Along with the main pipeline, the magnitude and center position of moving objects are measured automatically.

\section{Application in the YNHK survey}
\subsection{Detected asteroids}
The survey observations lasted for 6 years. With the special pipeline, we have reduced the 6 years of data of five fields of the YNHK survey. Here, we show the extraction result from the data of five selected fields obtained from April 2016 to June 2021. In total, there are 259 observing nights, and 24,188 images. The data reduction for asteroids is made based on all images of one field from each night. Finally, we have extracted a total of 538 lightcurves of 211 asteroids from the five fields. Table~\ref{tab1} lists numbers of asteroids, lightcurves, and detections in each of the five fields in columns 5--7. The columns 2--4 are the central positions of each field in equatorial coordinates and the number of observational nights for each field. 

\begin{table*}
    \caption{Asteroids identified in the YNHK survey.}
    \label{tab1}
    \centering
    \footnotesize% fontsize
    \setlength{\tabcolsep}{16pt}% column separation
    \renewcommand{\arraystretch}{1.0}%row space 
    \begin{threeparttable} 
    \begin{tabular}{cccccccccc}
\hline             
 Field name& RA & Dec & Observing & Number of &Number of&Number of\\
 & hh:mm:ss & dd:mm:ss & nights\tnote{1}\tnote& asteroids & lightcurves& detections\\
\hline
 May-Jun-01 & 14:35:20 & -15:20:00 & 101 (63)& 50& 121& 5744\\
 May-Jun-02 & 14:35:40 & -13:20:00 & 101 (54)& 52& 123& 5985\\
 May-Jun-03 & 14:41:20 & -17:40:00 & 101 (61)& 39& 96 & 4373\\
 May-Jun-04 & 14:58:00 & -19:00:00 & 96 (33)& 19& 49 & 2051\\
 Jan-Feb-01 & 06:24:00 & +28:00:00 & 189 (85)& 58& 149& 6249\\
\hline
    \end{tabular}
    \begin{tablenotes} 
        \item[1] Note: The number of nights that the asteroids were observed is in parentheses.
    \end{tablenotes}
    \end{threeparttable} 
\end{table*}

\begin{figure}
	\includegraphics[width=\columnwidth]{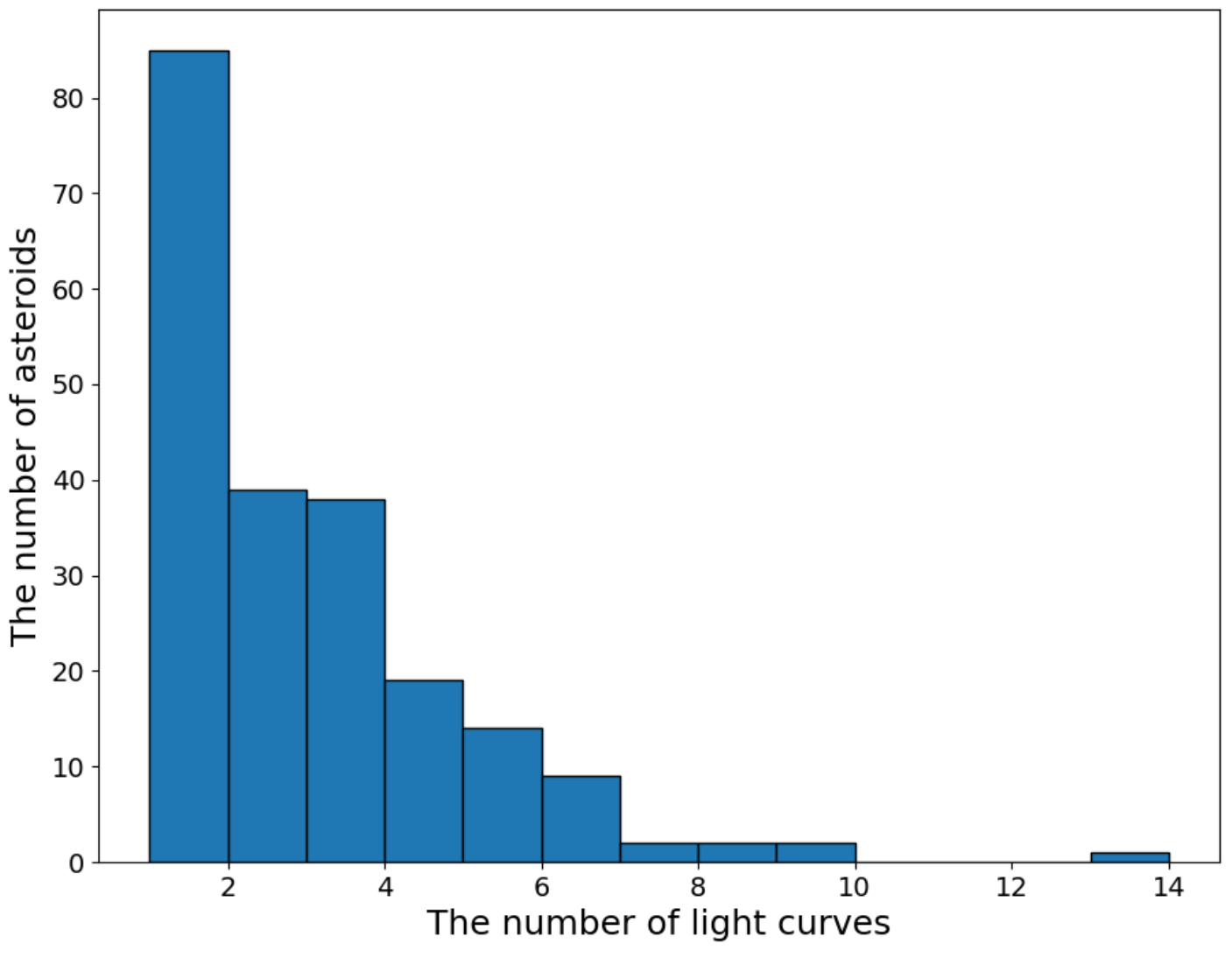}
    \caption{The number of lightcurves for detected asteroids in the YNHK survey.}
    \label{summary}
\end{figure}

The time spans of extracted lightcurves are roughly between 30 minutes and 7.8 hours, and the corresponding numbers of data points in individual lightcurve are between 8 and 95. A number of 507 lightcurves have more than 20 data points. We did statistics for the numbers of lightcurves for individual asteroids, presented in Fig. \ref{summary}. 85 \% of detected asteroids were observed for 1--4 nights, 11 \% were observed for 5--7 nights, and a few asteroids for 8--13 nights.

\subsection{Efficiency and accuracy of the identification} 

\begin{figure} 
\xdef\xfigwd{\textwidth}
    \centering
      \includegraphics[width=\columnwidth]{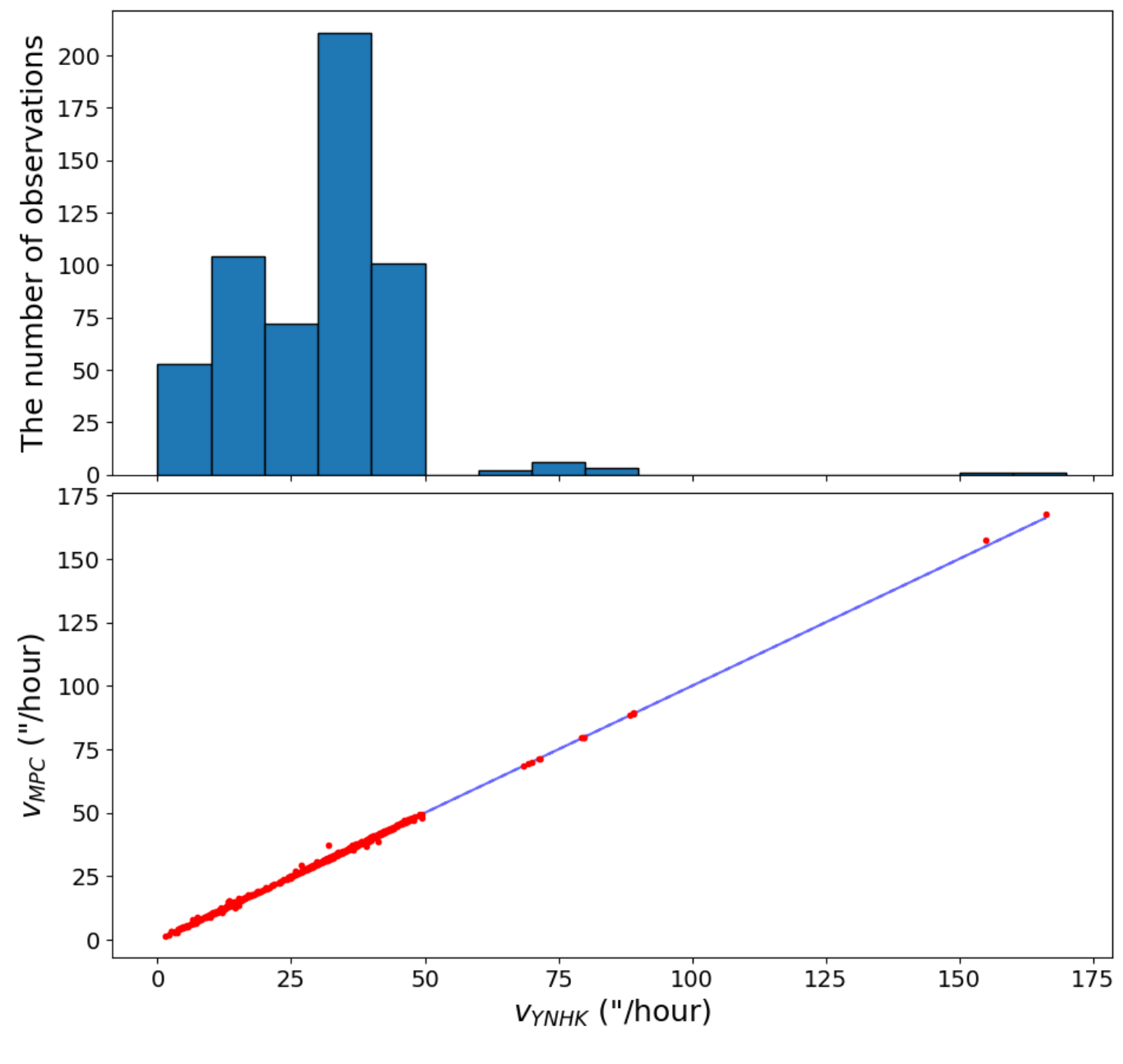}
	  \caption{Top: The histogram of the observed proper motions of 211 asteroids. Bottom: The comparison between the predicted and the observed proper motions of asteroids, the blue line represents the ideal relation.}
	  \label{motion} 
\end{figure}

Based on the detection results presented above, we would discuss the efficiency and accuracy of the methodology for the identification of SSOs in the YNHK survey. Among the detected moving objects, most are MBAs and a few are NEOs. The proper motions of detected asteroids derived by model trees are shown as a histogram plot in the upper panel of Fig. \ref{motion}. The proper motions of the detected asteroids range from 1".3 to 166" per hour, with the mode value at 27".12/hour. We compared the derived proper motions of 211 asteroids to that predicted by the MPC, and find a good agreement between them (see lower panel of Fig. \ref{motion}). The mean coordinate differences of SSO position between predicted by MPC service and measurements from our survey are -0".02 and -0".04 in RA and Dec. The corresponding standard deviations are 0".33 and 0".3208 (see Fig. \ref{dist}).

\begin{figure} 
\xdef\xfigwd{\textwidth}
    \centering
      \includegraphics[width=\columnwidth]{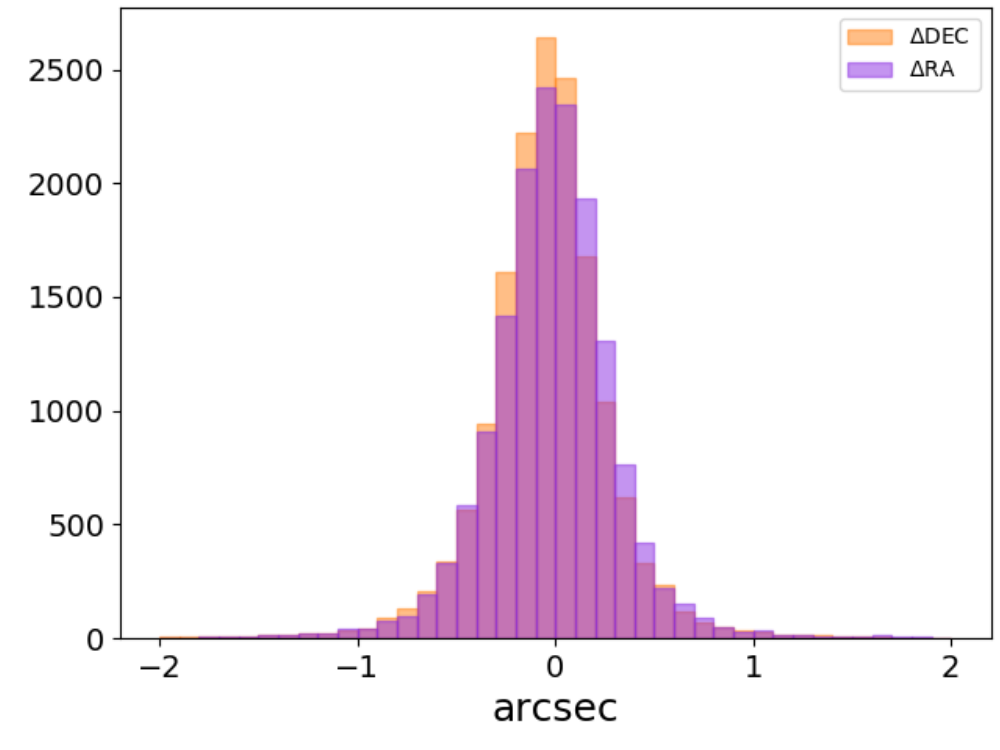}
	  \caption{Distribution of coordinate differences of asteroids between the predictions of MPC Checker service and the YNHK survey measurements.}
	  \label{dist} 
\end{figure}

\begin{table*}
    \caption{Asteroids' synodic rotational period derived from the YNHK survey observations.}
    \label{tab2}
    \centering
    \footnotesize% fontsize
    \setlength{\tabcolsep}{12pt}% column separation
    \renewcommand{\arraystretch}{1.0}%row space 
    \begin{tabular}{cccccccccc}
\hline             
Asteroid& Number of & Mag$_{V}$ & Phase angle & Derived  & Period (hour) & References \\
        &  nights &  &range (deg)& period (hour) & from the LCDB &\\
\hline
(320) Katharina    & 13& 16.4&11.4-16.4& 6.90$\pm$0.01  & 6.8926$\pm$0.0005  &Behrend 2006web\\
(379) Huenna       & 8 & 14.9&10.4-15.9& 7.07$\pm$0.01  & 14.1409$\pm$0.0003 &Behrend 2014web\\
(509) Iolanda      & 5 & 13.3&2.7-5.1  & 12.29$\pm$0.03 & 12.306$\pm$0.003   &\cite{Koff2000}\\
(1459) Magnya      & 3 & 15.6&2.3-2.9  & 4.68$\pm$0.03  & 4.678$\pm$0.001    &\cite{Licchelli2006}\\
(1841) Masaryk     & 4 & 15.5&2.6-3.7  & 7.53$\pm$0.01  & 7.53$\pm$0.04      &Behrend 2006web\\
(2007) McCuskey    & 3 & 15.0&11.3-12.8& 8.64$\pm$0.07  & 8.611$\pm$0.003    &\cite{Cantu2015}\\
(2572) Annschnell  & 3 & 16.1&9.9-11.4 & 6.30$\pm$0.05  & 6.343$\pm$0.003    &\cite{Stephens2017}\\
(2847) Parvati     & 6 & 14.5&1.6-5.2  & 2.64$\pm$0.01  & 2.6358$\pm$0.0001  &Pravec 2021web\\
(3458) Boduognat   & 3 & 16.6&22.3-22.8& 3.87$\pm$0.01  & 3.8565$\pm$0.0005  &\cite{Durkee2010}\\
(5278) Polly       & 4 & 16.9&19.0-22.9& 5.00$\pm$0.01  & -                  &-\\
(7999) Nesvorny    & 5 & 16.6&7.4-9.0  & 3.58$\pm$0.01  & -                  &-\\
(12915) 1998 SL161 & 2 & 16.3&6.2-6.7  & 10.60$\pm$0.05 & -                  &-\\
(26580) 2000 EW97  & 3 & 16.4&0.4-1.3  & 5.64$\pm$0.01  & -                  &-\\
\hline  
    \end{tabular}
\end{table*}

For the assessment of the detection efficiency of our methodology, as an example, we count the number of detected asteroids in the observations of the May-Jun-01 field during the 2016-2020 period, 50 asteroids are detected in all, among which the faintest asteroid has a magnitude of 17.45 in the V-band. Correspondingly, we check the predicted number of asteroids by the MPC service for the same observed field and date, and 71 asteroids with a limiting magnitude of 17.5 in the V band are predicted. That means we are not detecting 21 asteroids out of 71 in this field of the YNHK survey.

The reasons for the 21 asteroids lost are probably: 
(1) the limiting magnitude of the YNHK survey is 17.5 in the V-band according to stellar sources in a condition of clear weather, so the limiting magnitude will decrease when met cloudy weather; 
(2) because of a star tracking mode of the YNHK survey, the limiting magnitude of moving objects of the Solar System obviously decreases with their increasing proper motions; 
(3) those asteroids located at the edge of the field are difficult to detect due to the smaller number of detections in a single night's observing run. 
All of the 21 lost asteroids are fainter than 17 mag in the V band, and among them 18 asteroids are lost due to the bad weather, and 3 are close to the edge of field. 

\subsection{Determining synodic spin periods of the asteroids}

\begin{figure*} 
    \ContinuedFloat*
    \centering
        \includegraphics[width=0.98\linewidth]{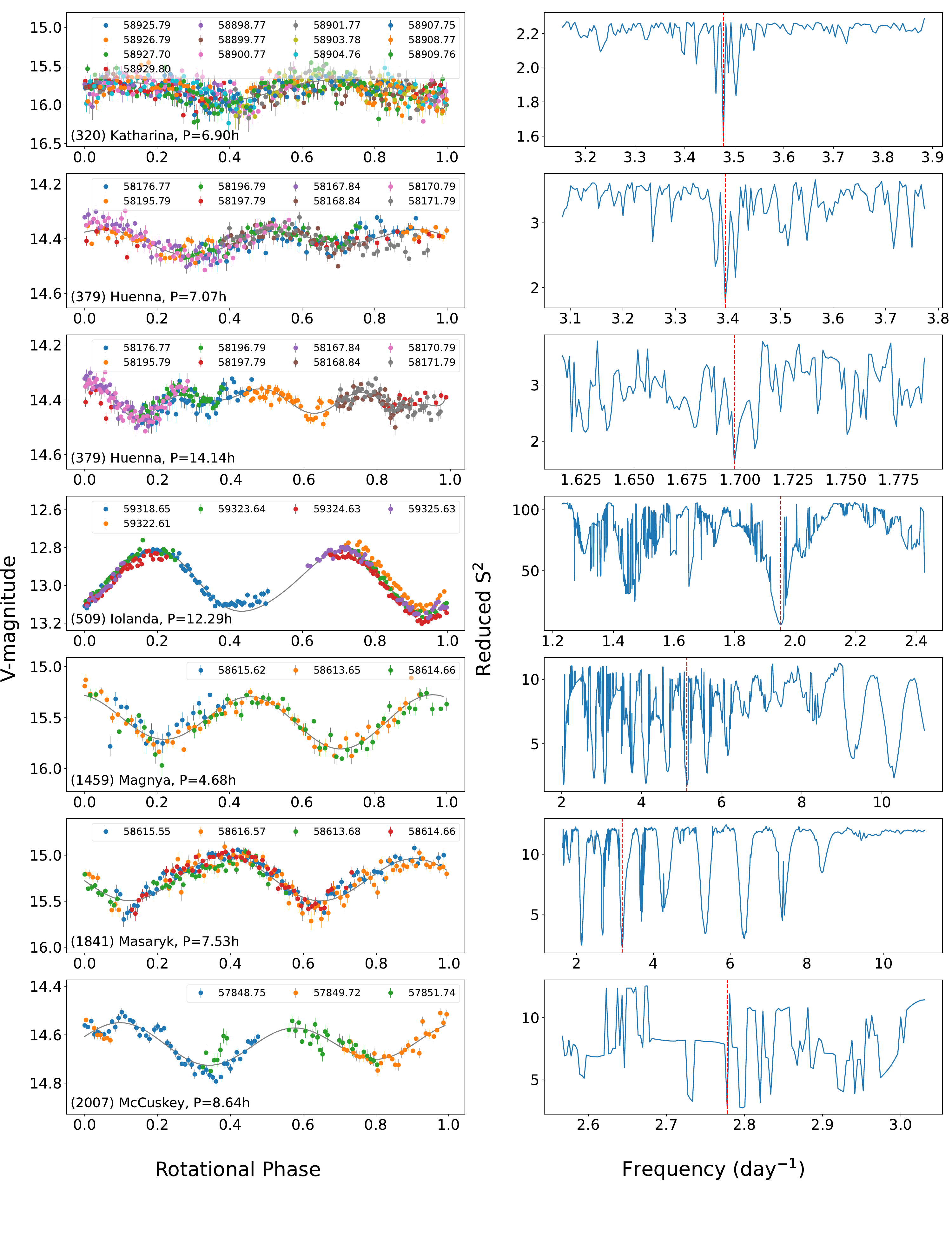}
	\label{lc}
	  \caption{Results of period analysis for 13 asteroids. Left column: Composite lightcurves folded with the best-fitting period, the legend of each symbol is MJD of lightcurve. Right column: Periodogram and the red dotted line indicates the best-fitted frequency.}
\end{figure*}

\begin{figure*} 
    \ContinuedFloat
    \centering
        \includegraphics[width=0.98\linewidth]{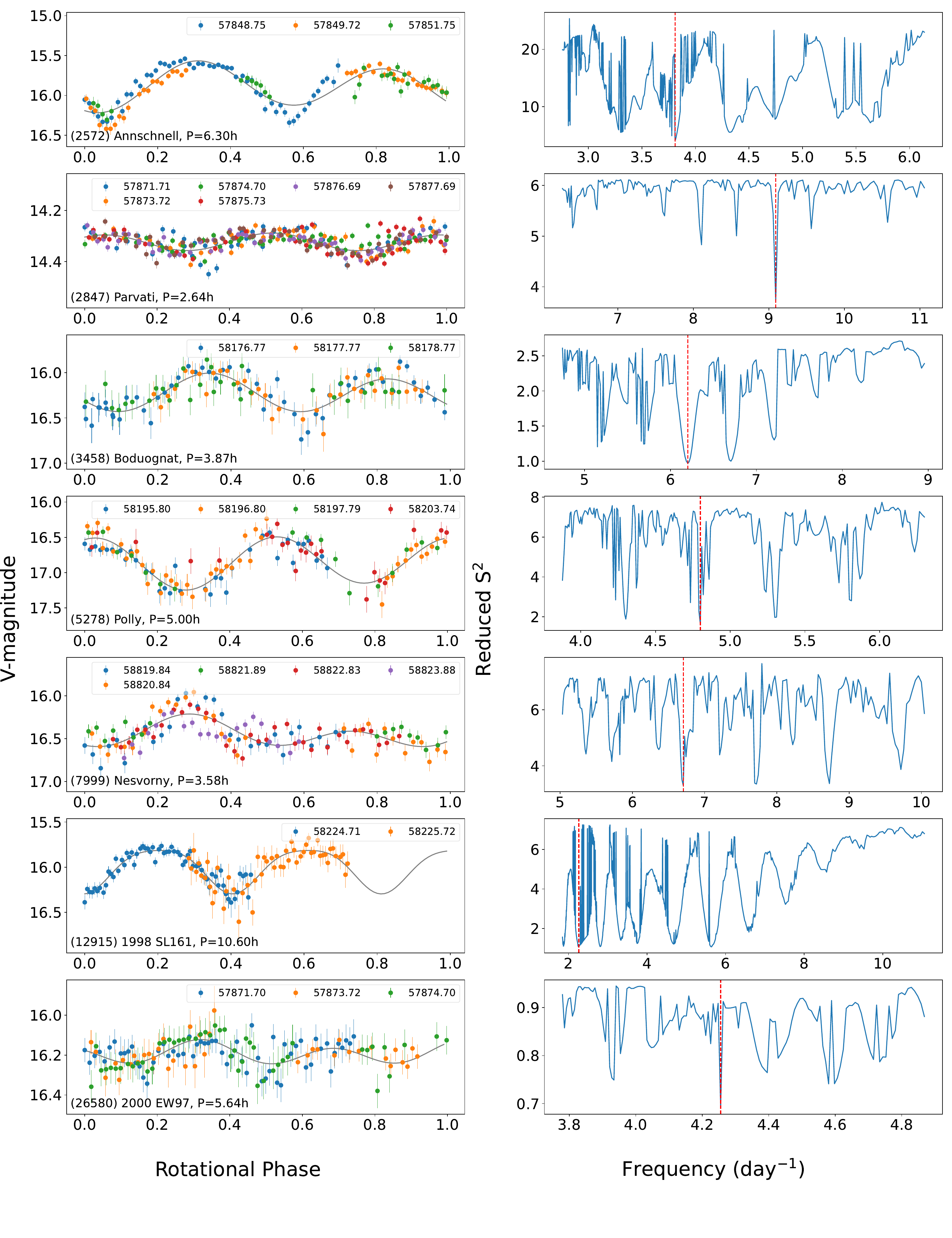}
	  \caption{(Continue)} 
\end{figure*}

From the YNHK survey, we can obtain multiple nights' lightcurves for many asteroids. The physical studies for these asteroids can be performed if combined with the existing data. We can determine the synodic spin periods of asteroids with our own data, which is especially valuable for those asteroids without previous spin information. We apply here the Fourier analysis method proposed by \cite{harris1989photoelectric} to determine the spin periods of 13 asteroids only using the lightcurves from the YNHK survey. The calibrated magnitude of an asteroid is modeled in the V-band with
\begin{equation}
   M(\alpha,t)=\Bar{M}(\alpha)+\sum_{l=1}^{\emph{mf}}A_{l}\sin\left[\frac{2{\pi}l}{P}(t-t_{0})\right]+B_{l}\cos\left[\frac{2{\pi}l}{P}(t-t_{0})\right]
	\label{eq:peroid}
\end{equation}
at phase angle $\alpha$ and time $t$, $\Bar{M}(\alpha)$ is the mean magnitude in individual night, and $t_{0}$ is the zero point of epoch. In Eq.~(\ref{eq:peroid}), the parameters involved are the Fourier coefficients $A_{l}$ and $B_{l}$, the spin period of asteroid $P$, and the mean magnitude in each observational night $\Bar{M}(\alpha)$. We use the least-squares method to derive these parameters by minimizing the reduced variance $S^2$ of Eq.~(\ref{eq:Chisq}), in which $k$ means the number of parameters and $\epsilon_i$ is the observational error. As for the degree of Fourier series, we set here $\emph{mf}=2$. For finding the global minimum of variance $S^2$, we scan the range of periods from 0.5 to 24 hours with a step of 0.01 hours. The variance is
\begin{equation}
    S^2=\frac{1}{n-k}\sum_{i=1}^{n}(\frac{V_i(\alpha_j,t_i)-M(\alpha_j,t_i)}{\epsilon_i})^2 \ ,
	\label{eq:Chisq}
\end{equation}
where $n$ is the total number of observations of an asteroid, $k=2 \emph{mf}+N_{d}+1$ is the degrees of freedom of the solution, $\emph{mf}$ is the degree of Fourier series, and $N_{d}$ is the number of days of observations. As for the uncertainty of rotation period, we applied the same method of \cite{harris1989photoelectric}, the uncertainty estimated by using the quantities of $S^2_{min}$ and $\Delta S^2$.

The period analysis results for the 13 asteroids are summarized in Table~\ref{tab2}, the composite lightcurves and periodograms are shown in Fig. \ref{lc}. Among these 13 asteroids, 9 have known spin periods, and for 4 the spin periods are derived for the first time. For the 9 asteroids, we compare our results to that listed in the LCDB, and, except for (379) Huenna, the derived spin periods are similar. The new derived period of (379) Huenna is half of the previous value in the LCDB. The possible reason is that the harmonics of the true period affect our solution. We have determined the periods of (5278) Polly, (7999) Nesvorny, (12915) 1998 SL161, and (26580) 2000 EW97 for the first time.

\textbf{(320) Katharina} was observed by the YNHK survey in 13 nights between 2020 February 19 and 2020 March 21. 
The solar phase angle of (320) Katharina ranged from $11.4^{\circ}$ to $16.4^{\circ}$ during the observations. We have derived a period of 6.90$\pm$0.01 hours, which is very close to the results of \cite{2012ApJ...759L...8M} and \cite{2015AJ....150...75W} in the LCDB. The amplitude of its composite lightcurve is around 0.25 mag. 
%mag/rad.

\textbf{(379) Huenna} passed one observed field of the YNHK survey during 2018 February 18 to 2018 March 20, we extract its data in 8 nights of observations. The solar phase angle during the observations of (379) Huenna ranged from $10.4^{\circ}$ to $15.9^{\circ}$. The Fourier analysis method gives a period of 7.07$\pm$0.01 hours with the minimum of reduced $S^2$. This result is consistent with the value of 7.002 hours of \cite{Harris1992}. In literature, Asteroid (379) Huenna was suggested as an asynchronous binary, \cite{Warner2010} estimated a period of 14.141 hours. We also search the period around 14.14 hours, and found a minimum of reduced $S^2$ at 1.89, which is slightly larger than that of the period of 7.07 hours (see Fig. \ref{lc}).
%mag/rad.

\textbf{(509) Iolanda}, an S-type asteroid, was observed by the YNHK survey in 5 nights during the period April 11-21,2021. The corresponding solar phase angle is from $5.1^{\circ}$ to $2.8^{\circ}$. The calibrated mean magnitudes of (509) are 8.537, 8.634, 8.540, 8.527, and 8.414 mag respectively, corresponding to phase angles $5.11^{\circ}$, $3.79^{\circ}$, $3.42^{\circ}$, $3.07^{\circ}$, and $2.72^{\circ}$. Such photometric data are important to study the opposition effect. Based on the five night data, we have derived a period value of 12.29$\pm$0.03 hours with an amplitude of 0.16 mag, which is very close to the result of \cite{Koff2000}, 12.306 hours, and slightly different from that of \cite{Lopez-Gonzalez2000}, 12.72 hours. \cite{Blanco2000} provided a longer period of 16.592 hours. Compared to published lightcurves, the peak to peak amplitude of our lightcurve is the smallest, 0.32 mag.

\textbf{(1459) Magnya} was caught by the YNHK survey in three continuous nights of May 10--12, 2019 when it was near opposition (solar phase angle of $2.3^{\circ}$--$2.9^{\circ}$). An unambiguous period of 4.68$\pm$0.03 hours is re-determined based on our data which is consistent with the previously published values \citep{Almeida2004,Behrendweb,Delbo2006,Licchelli2006,Silva2015,Durech2016,Hanus2016}. The amplitude of the composite lightcurve is around 0.58 mag. From the shape of the composite lightcurve, the asteroid may have a regular ellipsoidal shape. The average calibrated magnitude over the three nights is 10.31 mag.

\textbf{(1841) Masaryk} was also observed by the YNHK survey in four continuous nights of May 10--13, 2019, at its opposition period (solar phase angle of $2.6^{\circ}$--$3.7^{\circ}$). The derived period from our data is 7.53$\pm$0.01 hours with an amplitude of 0.52 mag, which is the same as other results \citep{Behrendweb,Durech2016}. Similar to asteroid (1459) Magnya, we have calculated the average calibrated magnitude over the four nights: it is 10.85 mag, close to its absolute magnitude of 10.8 mag.

\textbf{(2007) McCuskey}'s spin period is re-determined using 3 lightcurves obtained on April 4, 6, and 8, 2017, the derived period of 8.64$\pm$0.07 hours is consistent with the literature periods of \cite{Cantu2015} and \cite{Klinglesmith2013}. The amplitude of the composite lightcurve folded with the derived period is 0.13 mag.

\textbf{(2572) Annschnell} was detected simultaneously in the same field as (2007) McCuskey. The three night observations for (2572) Annschnell give a period of 6.30$\pm$0.05 hours and an amplitude of 0.55 mag. The re-determined period is close to those previous results \citep{Behrendweb,Stephens2017}, but with the smaller amplitude compared to the result of \cite{Behrendweb}.

\textbf{(2847) Parvati} was observed in 6 nights from April 28 to May 4 of 2017. In spite of the low amplitude of each lightcurve, an unambiguous period of 2.64$\pm$0.01 hours can be distinguished from the plot of frequency vs reduced $S^2$, and consistent with previous result \citep{Mas2018}. The amplitude of lightcurve is around 0.07 mag and the solar phase angle of this apparition observation ranges from $1.6^{\circ}$ to $5.2^{\circ}$. We have calculated the averaged calibrated magnitude of this asteroid as 12.95 mag, which is slightly fainter than the value of $H = 12.84$ mag in the MPC. 

\textbf{(3458) Boduognat} was observed on February 27-28 and March 1, 2018. The re-calculated period of this asteroid is 3.87$\pm$0.01 hours with an amplitude of 0.39 mag, which is slightly longer than that from \cite{Durkee2010}. From the shape of the lightcurve, (3458) Boduognat could be a regular ellipsoidal asteroid. The solar phase angle of the three nights ranges from $22.3^{\circ}$ to $22.8^{\circ}$, the averaged calibrated magnitude of the asteroid in this apparition is 13.52 mag.

\begin{figure*}
    \begin{minipage}{0.5\linewidth}
        \centering
        \includegraphics[width=0.8\textwidth]{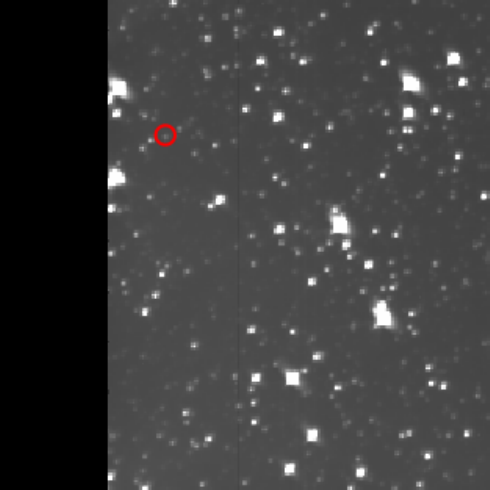}
    \end{minipage}%
    \begin{minipage}{0.5\linewidth}
        \centering
        \includegraphics[width=\textwidth]{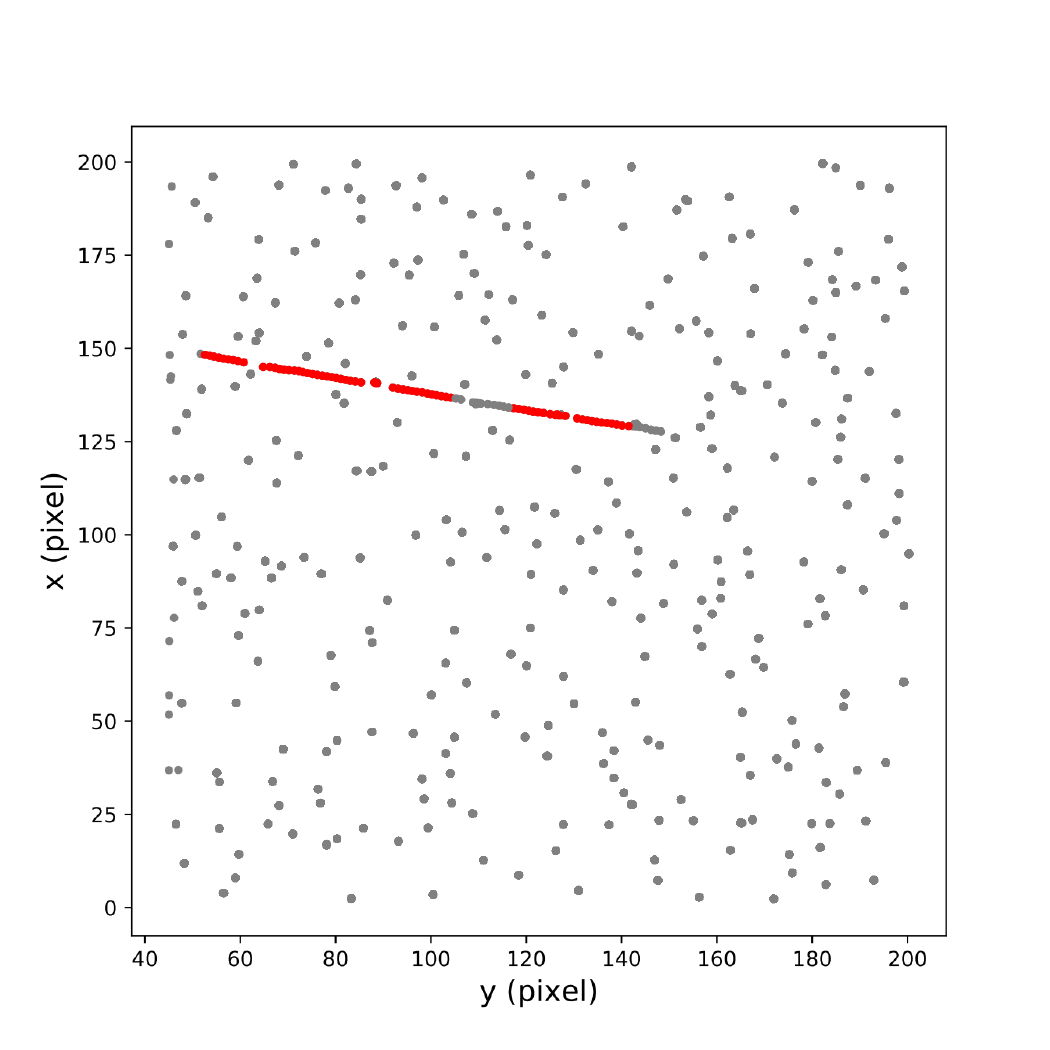}
    \end{minipage}
    \caption{TESS original image is shown in the left panel, the position of (1036) Ganymed is marked by a red circle. The reduced work dataset is shown in the right panel and red dots are tracklets of the asteroid identified by our method.}
    \label{TESS}
\end{figure*}

\textbf{(5278) Polly} has no spin information till now. We have determined its spin period with four nights of photometric data extracted from the YNHK survey, and derived a period of 5.00$\pm$0.01 hours with an amplitude of 0.70 mag. The solar phase angle of the observations varies from $19.0^{\circ}$ to $22.9^{\circ}$, the average calibrated magnitude of the asteroid in this apparition is 14.73 mag. 

\textbf{(7999) Nesvorny} was observed in 5 continuous nights of  Dec. 02--06, 2019. For the first time, we have derived a period of 3.58$\pm$0.01 hours with a low amplitude of 0.26 mag for this asteroid. The solar phase angle of the 5 night observations ranges from $7.4^{\circ}$ to $9.0^{\circ}$, the averaged calibrated magnitude of the asteroid in this apparition is 12.89 mag.

\textbf{(12915) 1998 SL161} has no earlier spin information, it was observed in 2 continuous nights of April 16 and 17, 2018. We have derived a period of 10.60$\pm$0.05 hours with an amplitude of 0.48 mag. From the composite lightcurve, two nights' data did not cover the whole rotational phase, but the overlap of two nights' data is significantly good. The solar phase angles for the 2 nights of observations are $6.7^{\circ}$ and $6.2^{\circ}$, the averaged calibrated magnitude of the asteroid in this apparition is 13.83 mag. 

\textbf{(26580) 2000 EW97} has 3 continuous nights of photometric data obtained by the YNHK survey from April 28 to May 01, 2017. The amplitude of its lightcurve is small, around 0.1 mag. With the Fourier analysis method, we have derived a spin period of 5.64$\pm$0.01 hours. The solar phase angles of the 3 nights of observations are small, from $0.4^{\circ}$ to $1.3^{\circ}$, its average calibrated magnitude in this apparition is 13.93 mag. This value near the opposition time is close to the MPC absolute magnitude ($H = 13.92$ mag).

\section{Discussion}

Our methodology can be applied to other time-domain surveys. For example, we have applied the methodology to the data of TESS between 2019 April 06 and 2019 April 08. The corresponding text images and the work dataset are made by trimming the full frame images of TESS into one square degree, from which one NEO, (1036) Ganymed is identified (see Fig. \ref{TESS}).

Because our methodology works on text images, it is easy to be used to search for asteroids with the data obtained by different telescopes at close dates, and for Kuiper belt objects (KBOs) over a long time span, as long as the data satisfy the detectable conditions of the model tree algorithm.

The cadences of data and the proper motion rates of moving objects determine whether they are detected with the model tree algorithm. The cadences of time-domain surveys vary according to their own scientific goals. For example, the data of TESS that we used to test our methodology mentioned above have a 30-min cadence. 

For searching moving objects of Solar System from a time-domain survey with our methodology, it is necessary to consider seriously the cadence of data where classes of small SSOs to be detected. Provided that moving objects, e.g., asteroids, are detectable in astronomical images of a given survey, the shorter the cadence of the data is, the more easily are the tracklets of asteroids identified.

However, if the proper motions of moving objects are too large and/or the cadence of data too long, two continuous detections of a moving object will be at large distances. Under such a situation, the model tree algorithm has difficulties in deriving any leaf node with a reasonable tracklet, because the node covers a large area and contains more noise data, i.e., non-moving sources in the work dataset. Probably, the node is divided further, and therefore this moving object is missed. For the case of 5-min cadence in the YNHK survey, we have made the simulation for testing the maximum proper motion of moving object where our methodology is still valid, and found the maximum distance interval to be 80" for two adjacent detections of a moving object, corresponding to a proper motion of 960"/hr. This value allows the identification for most moving objects of the Solar System, even some debris around the Earth. For detecting these lost faint asteroids in the YNHK survey fields, previously the cadence of 8 seconds was used but we plan to adopt longer cadence data, e.g., 2 min.

The computing time for a single night's data depends on a variety of factors, such as the number of images, the number of sources remaining after removing stellar sources in each text image, the threshold criteria TolN and TolS, and finally the computer hardware. As a reference, a complete analysis of a work dataset with 2,116 detections takes 23 s when using an Intel Core i5-9500 computer with 8 GB RAM. In the future, we shall consider parallel processing to speed up the computation. For the most important feature of our methodology, the portability of the algorithm to other surveys, we are going to apply it to mine the data of small SSOs from the future Chinese Space Station Telescope (CSST) survey.

The CSST is going to survey 17,500 square degree sky areas in its ten year operation, in which numerous of small objects of the Solar System with apparent magnitude brighter than 26 mag in the r-band will be contained in the survey images. Assuming its 150 s exposure of survey observations and preprogrammed observation strategy of the CSST survey, at least the moving objects in far outer space of the Solar System, e.g., transneptunian objects (TNOs) and KBOs, could be detected with our methodology. We believe that the understanding for the small objects of Solar System will be improved significantly through deriving their physical characteristics based on the future CSST's observations.

\section{Summary}
The physical parameters of asteroids are important for understanding the origin and evolution of these objects. Time-series photometric data are a key to getting the physical parameters of asteroids. Here, we have developed a machine learning based methodology to identify asteroids and to extract their photometric data from time-domain surveys. We conclude that:

(1) The methodology works on text images and is fast and flexible to be applied to different time-domain surveys.

(2) We have tested this method with the data from five fields of the YNHK survey, from which 538 lightcurves of 211 asteroids are extracted. Among the 211 asteroids, 40 \% of them have only 1 lightcurve, and 45 \% of them have 2--4 lightcurves, 11 \% of them have 5--7 lightcurves, the rest 4 \% have 8--13 lightcurves. Such data will contribute to the photometric inversion of these asteroids. At least, we can determine the spin periods of asteroids with these extracted lightcurves. As an example, we derive the results of spin determination for 13 asteroids, among which 4 ones are newly derived. Other sky areas of the YNHK survey are under analysis.

(3) The method suits for detecting all small objects of the Solar System according to the present simulation tests as long as there are more than six detections for moving objects in the work dataset and their proper motions are slower than 960"/hour.

(4) We are going to apply the method to search for outer small objects of the Solar System, e.g., KBOs, Centaurs, TNOs, from the data produced by the future CSST survey.

\section*{Acknowledgements}
This work is supported by the National Natural Science Foundation of China (grant Nos. 11073051, U1531121, 11673063 and 12003063) and the Academy of Finland (grants No. 325805, 336546, and 345115). We acknowledge the science research grants from the China Manned Space Project with No. CMS-CSST-2021-B08 and the Chinese Academy of Sciences President’s International Fellowship Initiative (PIFI) Grant No. 2021VMA0017. This work has made use of the Minor Planet Checker service provided by the Minor Planet Center. This paper includes data collected by the TESS mission. Funding for the TESS mission is provided by the NASA’s Science Mission Directorate. This research has made use of the VizieR catalogue access tool, CDS, Strasbourg, France. This work has made use of data from the European Space Agency mission {\it Gaia} (\url{https://www.cosmos.esa.int/gaia}), processed by the {\it Gaia} Data Processing and Analysis Consortium (DPAC, \url{https://www.cosmos.esa.int/web/gaia/dpac/consortium}). Funding for the DPAC has been provided by national institutions, in particular the institutions participating in the {\it Gaia} Multilateral Agreement.

\section*{Data availability}
The TESS data underlying this article are available from the Mikulski Archive for Space Telescopes (MAST) at \url{https://mast.stsci.edu}. The YNHK survey data underlying this article can be shared by contacting to the corresponding author. 

\bibliographystyle{mnras}
\bibliography{mnras} % if your bibtex file is called example.bib

\begin{thebibliography}{}
\makeatletter
\relax
\def\mn@urlcharsother{\let\do\@makeother \do\$\do\&\do\#\do\^\do\_\do\%\do\~}
\def\mn@doi{\begingroup\mn@urlcharsother \@ifnextchar [ {\mn@doi@}
  {\mn@doi@[]}}
\def\mn@doi@[#1]#2{\def\@tempa{#1}\ifx\@tempa\@empty \href
  {http://dx.doi.org/#2} {doi:#2}\else \href {http://dx.doi.org/#2} {#1}\fi
  \endgroup}
\def\mn@eprint#1#2{\mn@eprint@#1:#2::\@nil}
\def\mn@eprint@arXiv#1{\href {http://arxiv.org/abs/#1} {{\tt arXiv:#1}}}
\def\mn@eprint@dblp#1{\href {http://dblp.uni-trier.de/rec/bibtex/#1.xml}
  {dblp:#1}}
\def\mn@eprint@#1:#2:#3:#4\@nil{\def\@tempa {#1}\def\@tempb {#2}\def\@tempc
  {#3}\ifx \@tempc \@empty \let \@tempc \@tempb \let \@tempb \@tempa \fi \ifx
  \@tempb \@empty \def\@tempb {arXiv}\fi \@ifundefined
  {mn@eprint@\@tempb}{\@tempb:\@tempc}{\expandafter \expandafter \csname
  mn@eprint@\@tempb\endcsname \expandafter{\@tempc}}}

\bibitem[\protect\citeauthoryear{{Almeida}, {Angeli}, {Duffard}  \&
  {Lazzaro}}{{Almeida} et~al.}{2004}]{Almeida2004}
{Almeida} R.,  {Angeli} C.~A.,  {Duffard} R.,   {Lazzaro} D.,  2004, \mn@doi
  [\aap] {10.1051/0004-6361:20034585}, \href
  {https://ui.adsabs.harvard.edu/abs/2004A&A...415..403A} {415, 403}

\bibitem[\protect\citeauthoryear{{Bakos}, {Noyes}, {Kov{\'a}cs}, {Stanek},
  {Sasselov}  \& {Domsa}}{{Bakos} et~al.}{2004}]{HATNet}
{Bakos} G.,  {Noyes} R.~W.,  {Kov{\'a}cs} G.,  {Stanek} K.~Z.,  {Sasselov}
  D.~D.,   {Domsa} I.,  2004, \mn@doi [\pasp] {10.1086/382735}, \href
  {https://ui.adsabs.harvard.edu/abs/2004PASP..116..266B} {116, 266}

\bibitem[\protect\citeauthoryear{Behrend}{Behrend}{2006}]{Behrendweb}
Behrend R.,  2005,2006, Observatoire de Geneve web site,
  \url{http://obswww.unige.ch/~behrend/page_cou.html}

\bibitem[\protect\citeauthoryear{{Blanco}, {Di Martino}  \&
  {Riccioli}}{{Blanco} et~al.}{2000}]{Blanco2000}
{Blanco} C.,  {Di Martino} M.,   {Riccioli} D.,  2000, \mn@doi [\planss]
  {10.1016/S0032-0633(99)00074-4}, \href
  {https://ui.adsabs.harvard.edu/abs/2000P&SS...48..271B} {48, 271}

\bibitem[\protect\citeauthoryear{{Borucki} et~al.,}{{Borucki}
  et~al.}{2010}]{Kepler}
{Borucki} W.~J.,  et~al., 2010, \mn@doi [Science] {10.1126/science.1185402},
  \href {https://ui.adsabs.harvard.edu/abs/2010Sci...327..977B} {327, 977}

\bibitem[\protect\citeauthoryear{{Bowell}, {Koehn}, {Howell}, {Hoffman}  \&
  {Muinonen}}{{Bowell} et~al.}{1995}]{LONEOS}
{Bowell} E.,  {Koehn} B.~W.,  {Howell} S.~B.,  {Hoffman} M.,   {Muinonen} K.,
  1995, in AAS/Division for Planetary Sciences Meeting Abstracts \#27. p. 01.10

\bibitem[\protect\citeauthoryear{Breiman, Friedman, Stone  \& Olshen}{Breiman
  et~al.}{1984}]{breiman1984classification}
Breiman L.,  Friedman J.,  Stone C.,   Olshen R.,  1984, Classification and
  Regression Trees.
Taylor \& Francis

\bibitem[\protect\citeauthoryear{{Broeg, C.} et~al.,}{{Broeg, C.}
  et~al.}{2013}]{CHEOPS}
{Broeg, C.} et~al., 2013, \mn@doi [EPJ Web of Conferences]
  {10.1051/epjconf/20134703005}, 47, 03005

\bibitem[\protect\citeauthoryear{{Cantu}, {Adolphson}, {Montgomery}  \&
  {Renshaw}}{{Cantu} et~al.}{2015}]{Cantu2015}
{Cantu} S.,  {Adolphson} M.,  {Montgomery} K.,   {Renshaw} T.,  2015, Minor
  Planet Bulletin, \href
  {https://ui.adsabs.harvard.edu/abs/2015MPBu...42...28C} {42, 28}

\bibitem[\protect\citeauthoryear{{Collier Cameron} et~al.,}{{Collier Cameron}
  et~al.}{2006}]{collier2006fast}
{Collier Cameron} A.,  et~al., 2006, \mn@doi [\mnras]
  {10.1111/j.1365-2966.2006.11074.x}, \href
  {https://ui.adsabs.harvard.edu/abs/2006MNRAS.373..799C} {373, 799}

\bibitem[\protect\citeauthoryear{{Cort{\'e}s-Contreras}
  et~al.,}{{Cort{\'e}s-Contreras} et~al.}{2019}]{10.1093/mnras/stz2727}
{Cort{\'e}s-Contreras} M.,  et~al., 2019, \mn@doi [\mnras]
  {10.1093/mnras/stz2727}, \href
  {https://ui.adsabs.harvard.edu/abs/2019MNRAS.490.3046C} {490, 3046}

\bibitem[\protect\citeauthoryear{{Delbo} et~al.,}{{Delbo}
  et~al.}{2006}]{Delbo2006}
{Delbo} M.,  et~al., 2006, \mn@doi [\icarus] {10.1016/j.icarus.2006.01.001},
  \href {https://ui.adsabs.harvard.edu/abs/2006Icar..181..618D} {181, 618}

\bibitem[\protect\citeauthoryear{{Denneau} et~al.,}{{Denneau}
  et~al.}{2013}]{Pan-STARRS-System}
{Denneau} L.,  et~al., 2013, \mn@doi [\pasp] {10.1086/670337}, \href
  {https://ui.adsabs.harvard.edu/abs/2013PASP..125..357D} {125, 357}

\bibitem[\protect\citeauthoryear{{Durech}, {Sidorin}  \&
  {Kaasalainen}}{{Durech} et~al.}{2010}]{2010A/&A...513A..46D}
{Durech} J.,  {Sidorin} V.,   {Kaasalainen} M.,  2010, \mn@doi [\aap]
  {10.1051/0004-6361/200912693}, \href
  {https://ui.adsabs.harvard.edu/abs/2010A&A...513A..46D} {513, A46}

\bibitem[\protect\citeauthoryear{{Durkee}}{{Durkee}}{2010}]{Durkee2010}
{Durkee} R.~I.,  2010, Minor Planet Bulletin, \href
  {https://ui.adsabs.harvard.edu/abs/2010MPBu...37..125D} {37, 125}

\bibitem[\protect\citeauthoryear{{Erasmus}, {McNeill}, {Mommert}, {Trilling},
  {Sickafoose}  \& {Paterson}}{{Erasmus} et~al.}{2019}]{Erasmus2019}
{Erasmus} N.,  {McNeill} A.,  {Mommert} M.,  {Trilling} D.~E.,  {Sickafoose}
  A.~A.,   {Paterson} K.,  2019, \mn@doi [\apjs] {10.3847/1538-4365/ab1344},
  \href {https://ui.adsabs.harvard.edu/abs/2019ApJS..242...15E} {242, 15}

\bibitem[\protect\citeauthoryear{{Gaia Collaboration} et~al.,}{{Gaia
  Collaboration} et~al.}{2018}]{spoto2018gaia}
{Gaia Collaboration} et~al., 2018, \mn@doi [\aap]
  {10.1051/0004-6361/201832900}, \href
  {https://ui.adsabs.harvard.edu/abs/2018A&A...616A..13G} {616, A13}

\bibitem[\protect\citeauthoryear{{Gaia Collaboration} et~al.,}{{Gaia
  Collaboration} et~al.}{2021}]{GaiaEDR32021}
{Gaia Collaboration} et~al., 2021, \mn@doi [\aap]
  {10.1051/0004-6361/202039657}, \href
  {https://ui.adsabs.harvard.edu/abs/2021A&A...649A...1G} {649, A1}

\bibitem[\protect\citeauthoryear{Gu et~al.,}{Gu et~al.}{2022}]{gu2022}
Gu S.,  et~al., 2022, \mn@doi [Astronomische Nachrichten]
  {https://doi.org/10.1002/asna.20224022}, 343, e20224022

\bibitem[\protect\citeauthoryear{{Hanu{\v{s}}} et~al.,}{{Hanu{\v{s}}}
  et~al.}{2016}]{Hanus2016}
{Hanu{\v{s}}} J.,  et~al., 2016, \mn@doi [\aap] {10.1051/0004-6361/201527441},
  \href {https://ui.adsabs.harvard.edu/abs/2016A&A...586A.108H} {586, A108}

\bibitem[\protect\citeauthoryear{Harrington}{Harrington}{2012}]{10.5555/2361796}
Harrington P.,  2012, Machine Learning in Action.
Manning Publications Co., USA

\bibitem[\protect\citeauthoryear{{Harris} et~al.,}{{Harris}
  et~al.}{1989}]{harris1989photoelectric}
{Harris} A.~W.,  et~al., 1989, \mn@doi [\icarus]
  {10.1016/0019-1035(89)90015-8}, \href
  {https://ui.adsabs.harvard.edu/abs/1989Icar...77..171H} {77, 171}

\bibitem[\protect\citeauthoryear{{Harris}, {Young}, {Dockweiler}, {Gibson},
  {Poutanen}  \& {Bowell}}{{Harris} et~al.}{1992}]{Harris1992}
{Harris} A.~W.,  {Young} J.~W.,  {Dockweiler} T.,  {Gibson} J.,  {Poutanen} M.,
    {Bowell} E.,  1992, \mn@doi [\icarus] {10.1016/0019-1035(92)90195-D}, \href
  {https://ui.adsabs.harvard.edu/abs/1992Icar...95..115H} {95, 115}

\bibitem[\protect\citeauthoryear{{Ivezi{\'c}} et~al.,}{{Ivezi{\'c}}
  et~al.}{2001}]{Ivezi2001}
{Ivezi{\'c}} {\v{Z}}.,  et~al., 2001, \mn@doi [\aj] {10.1086/323452}, \href
  {https://ui.adsabs.harvard.edu/abs/2001AJ....122.2749I} {122, 2749}

\bibitem[\protect\citeauthoryear{{Jones} et~al.,}{{Jones} et~al.}{2018}]{LSST}
{Jones} R.~L.,  et~al., 2018, \mn@doi [\icarus] {10.1016/j.icarus.2017.11.033},
  \href {https://ui.adsabs.harvard.edu/abs/2018Icar..303..181J} {303, 181}

\bibitem[\protect\citeauthoryear{{Kaiser} et~al.,}{{Kaiser}
  et~al.}{2002}]{Pan-STARRS}
{Kaiser} N.,  et~al., 2002, in {Tyson} J.~A.,  {Wolff} S.,  eds,  Society of
  Photo-Optical Instrumentation Engineers (SPIE) Conference Series Vol. 4836,
  Survey and Other Telescope Technologies and Discoveries. pp 154--164,
  \mn@doi{10.1117/12.457365}

\bibitem[\protect\citeauthoryear{{Klinglesmith} \& {Franco}}{{Klinglesmith} \&
  {Franco}}{2013}]{Klinglesmith2013}
{Klinglesmith} Daniel~A. I.,  {Franco} Lorenzo F.,  2013, Minor Planet
  Bulletin, \href {https://ui.adsabs.harvard.edu/abs/2013MPBu...40..177K} {40,
  177}

\bibitem[\protect\citeauthoryear{{Koff} \& {Brincat}}{{Koff} \&
  {Brincat}}{2000}]{Koff2000}
{Koff} R.~A.,  {Brincat} S.~M.,  2000, Minor Planet Bulletin, \href
  {https://ui.adsabs.harvard.edu/abs/2000MPBu...27...49K} {27, 49}

\bibitem[\protect\citeauthoryear{{Kruk} et~al.,}{{Kruk}
  et~al.}{2022}]{Kruk2022}
{Kruk} S.,  et~al., 2022, in SciOps 2022: Artificial Intelligence for Science
  and Operations in Astronomy (SCIOPS). Proceedings of the ESA/ESO SCOPS
  Workshop held 16-20 May. p.~29, \mn@doi{10.5281/zenodo.6574489}

\bibitem[\protect\citeauthoryear{{Kubica}}{{Kubica}}{2005}]{2005PhDT........14K}
{Kubica} J.,  2005, PhD thesis, Carnegie Mellon University, Pennsylvania

\bibitem[\protect\citeauthoryear{{Lang}, {Hogg}, {Mierle}, {Blanton}  \&
  {Roweis}}{{Lang} et~al.}{2010}]{2010AJ....139.1782L}
{Lang} D.,  {Hogg} D.~W.,  {Mierle} K.,  {Blanton} M.,   {Roweis} S.,  2010,
  \mn@doi [\aj] {10.1088/0004-6256/139/5/1782}, \href
  {https://ui.adsabs.harvard.edu/abs/2010AJ....139.1782L} {139, 1782}

\bibitem[\protect\citeauthoryear{{Larson}, {Brownlee}, {Hergenrother}  \&
  {Spahr}}{{Larson} et~al.}{1998}]{CSS}
{Larson} S.,  {Brownlee} J.,  {Hergenrother} C.,   {Spahr} T.,  1998, in
  Bulletin of the American Astronomical Society. p.~1037

\bibitem[\protect\citeauthoryear{{Licchelli}}{{Licchelli}}{2006}]{Licchelli2006}
{Licchelli} D.,  2006, Minor Planet Bulletin, \href
  {https://ui.adsabs.harvard.edu/abs/2006MPBu...33...11L} {33, 11}

\bibitem[\protect\citeauthoryear{Liu, Ting  \& Zhou}{Liu
  et~al.}{2008}]{4781136}
Liu F.~T.,  Ting K.~M.,   Zhou Z.-H.,  2008, in Proceedings of the 2008 Eighth
  IEEE International Conference on Data Mining. ICDM '08.
IEEE Computer Society, USA, p. 413–422, \mn@doi{10.1109/ICDM.2008.17}, \url
  {https://doi.org/10.1109/ICDM.2008.17}

\bibitem[\protect\citeauthoryear{{Liu} et~al.,}{{Liu}
  et~al.}{2022}]{2022AJ....163..167L}
{Liu} Y.,  et~al., 2022, \mn@doi [\aj] {10.3847/1538-3881/ac50ab}, \href
  {https://ui.adsabs.harvard.edu/abs/2022AJ....163..167L} {163, 167}

\bibitem[\protect\citeauthoryear{{L{\'o}pez-Gonz{\'a}lez} \&
  {Rodr{\'\i}guez}}{{L{\'o}pez-Gonz{\'a}lez} \&
  {Rodr{\'\i}guez}}{2000}]{Lopez-Gonzalez2000}
{L{\'o}pez-Gonz{\'a}lez} M.~J.,  {Rodr{\'\i}guez} E.,  2000, \mn@doi [\aaps]
  {10.1051/aas:2000105}, \href
  {https://ui.adsabs.harvard.edu/abs/2000A&AS..145..255L} {145, 255}

\bibitem[\protect\citeauthoryear{{Luo} et~al.,}{{Luo}
  et~al.}{2022}]{10.1093/mnras/stac1406}
{Luo} X.,  et~al., 2022, \mn@doi [\mnras] {10.1093/mnras/stac1406}, \href
  {https://ui.adsabs.harvard.edu/abs/2022MNRAS.514.1511L} {514, 1511}

\bibitem[\protect\citeauthoryear{{Mainzer} et~al.,}{{Mainzer}
  et~al.}{2011}]{NEOWISE}
{Mainzer} A.,  et~al., 2011, \mn@doi [\apj] {10.1088/0004-637X/731/1/53}, \href
  {https://ui.adsabs.harvard.edu/abs/2011ApJ...731...53M} {731, 53}

\bibitem[\protect\citeauthoryear{{Martikainen}, {Muinonen}, {Penttil{\"a}},
  {Cellino}  \& {Wang}}{{Martikainen} et~al.}{2021}]{Martikainen2021}
{Martikainen} J.,  {Muinonen} K.,  {Penttil{\"a}} A.,  {Cellino} A.,   {Wang}
  X.~B.,  2021, \mn@doi [\aap] {10.1051/0004-6361/202039796}, \href
  {https://ui.adsabs.harvard.edu/abs/2021A&A...649A..98M} {649, A98}

\bibitem[\protect\citeauthoryear{{Mas} et~al.,}{{Mas} et~al.}{2018}]{Mas2018}
{Mas} V.,  et~al., 2018, Minor Planet Bulletin, \href
  {https://ui.adsabs.harvard.edu/abs/2018MPBu...45...76M} {45, 76}

\bibitem[\protect\citeauthoryear{{Masiero}, {Mainzer}, {Grav}, {Bauer},
  {Cutri}, {Nugent}  \& {Cabrera}}{{Masiero}
  et~al.}{2012}]{2012ApJ...759L...8M}
{Masiero} J.~R.,  {Mainzer} A.~K.,  {Grav} T.,  {Bauer} J.~M.,  {Cutri} R.~M.,
  {Nugent} C.,   {Cabrera} M.~S.,  2012, \mn@doi [\apjl]
  {10.1088/2041-8205/759/1/L8}, \href
  {https://ui.adsabs.harvard.edu/abs/2012ApJ...759L...8M} {759, L8}

\bibitem[\protect\citeauthoryear{{Michel}, {DeMeo}  \& {Bottke}}{{Michel}
  et~al.}{2015}]{AsteroidsIV}
{Michel} P.,  {DeMeo} F.~E.,   {Bottke} W.~F.,  2015, in , Asteroids IV.
University of Arizona Press, pp 3--10,
  \mn@doi{10.2458/azu_uapress_9780816532131-ch001}

\bibitem[\protect\citeauthoryear{{Mommert}}{{Mommert}}{2017}]{Mommert2017A&C}
{Mommert} M.,  2017, \mn@doi [Astronomy and Computing]
  {10.1016/j.ascom.2016.11.002}, \href
  {https://ui.adsabs.harvard.edu/abs/2017A&C....18...47M} {18, 47}

\bibitem[\protect\citeauthoryear{{Muinonen}, {Uvarova}, {Martikainen},
  {Penttil{\"a}}, {Cellino}  \& {Wang}}{{Muinonen} et~al.}{2022}]{Muinonen2022}
{Muinonen} K.,  {Uvarova} E.,  {Martikainen} J.,  {Penttil{\"a}} A.,  {Cellino}
  A.,   {Wang} X.~B.,  2022, \mn@doi [Front. Astron. Space Sci.]
  {10.3389/fspas.2022.821125}, 9, 1

\bibitem[\protect\citeauthoryear{{Pollacco} et~al.,}{{Pollacco}
  et~al.}{2006}]{WASP}
{Pollacco} D.~L.,  et~al., 2006, \mn@doi [\pasp] {10.1086/508556}, \href
  {https://ui.adsabs.harvard.edu/abs/2006PASP..118.1407P} {118, 1407}

\bibitem[\protect\citeauthoryear{{Popescu} et~al.,}{{Popescu}
  et~al.}{2016}]{Popescu2016}
{Popescu} M.,  et~al., 2016, \mn@doi [\aap] {10.1051/0004-6361/201628163},
  \href {https://ui.adsabs.harvard.edu/abs/2016A&A...591A.115P} {591, A115}

\bibitem[\protect\citeauthoryear{{Pravdo} et~al.,}{{Pravdo}
  et~al.}{1999}]{NEAT}
{Pravdo} S.~H.,  et~al., 1999, \mn@doi [\aj] {10.1086/300769}, \href
  {https://ui.adsabs.harvard.edu/abs/1999AJ....117.1616P} {117, 1616}

\bibitem[\protect\citeauthoryear{Quinlan}{Quinlan}{1986}]{quinlan1986}
Quinlan J.~R.,  1986, \mn@doi [Machine Learning] {10.1007/BF00116251}, 1, 81

\bibitem[\protect\citeauthoryear{Quinlan}{Quinlan}{1993}]{Quinlan1993}
Quinlan J.~R.,  1993, C4.5: Programs for Machine Learning.
Morgan Kaufmann, San Francisco (CA),
  \mn@doi{https://doi.org/10.1016/C2009-0-27846-9}

\bibitem[\protect\citeauthoryear{{Rabinowitz}}{{Rabinowitz}}{1991}]{Spacewatch}
{Rabinowitz} D.~L.,  1991, \mn@doi [\aj] {10.1086/115785}, \href
  {https://ui.adsabs.harvard.edu/abs/1991AJ....101.1518R} {101, 1518}

\bibitem[\protect\citeauthoryear{{Ricker} et~al.,}{{Ricker}
  et~al.}{2015}]{TESS}
{Ricker} G.~R.,  et~al., 2015, \mn@doi [Journal of Astronomical Telescopes,
  Instruments, and Systems] {10.1117/1.JATIS.1.1.014003}, \href
  {https://ui.adsabs.harvard.edu/abs/2015JATIS...1a4003R} {1, 014003}

\bibitem[\protect\citeauthoryear{{Sergeyev} \& {Carry}}{{Sergeyev} \&
  {Carry}}{2021}]{Sergeyev2021}
{Sergeyev} A.~V.,  {Carry} B.,  2021, \mn@doi [\aap]
  {10.1051/0004-6361/202140430}, \href
  {https://ui.adsabs.harvard.edu/abs/2021A&A...652A..59S} {652, A59}

\bibitem[\protect\citeauthoryear{{Sergeyev}, {Carry}, {Onken}, {Devillepoix},
  {Wolf}  \& {Chang}}{{Sergeyev} et~al.}{2022}]{Sergeyev2022}
{Sergeyev} A.~V.,  {Carry} B.,  {Onken} C.~A.,  {Devillepoix} H.~A.~R.,  {Wolf}
  C.,   {Chang} S.~W.,  2022, \mn@doi [\aap] {10.1051/0004-6361/202142074},
  \href {https://ui.adsabs.harvard.edu/abs/2022A&A...658A.109S} {658, A109}

\bibitem[\protect\citeauthoryear{{Silva} \& {Lazzaro}}{{Silva} \&
  {Lazzaro}}{2015}]{Silva2015}
{Silva} J.~S.,  {Lazzaro} D.,  2015, \mn@doi [\aap]
  {10.1051/0004-6361/201526350}, \href
  {https://ui.adsabs.harvard.edu/abs/2015A&A...580A..70S} {580, A70}

\bibitem[\protect\citeauthoryear{{Stephens}}{{Stephens}}{2017}]{Stephens2017}
{Stephens} R.~D.,  2017, Minor Planet Bulletin, \href
  {https://ui.adsabs.harvard.edu/abs/2017MPBu...44..321S} {44, 321}

\bibitem[\protect\citeauthoryear{{Stokes}, {Evans}, {Viggh}, {Shelly}  \&
  {Pearce}}{{Stokes} et~al.}{2000}]{LINEAR}
{Stokes} G.~H.,  {Evans} J.~B.,  {Viggh} H. E.~M.,  {Shelly} F.~C.,   {Pearce}
  E.~C.,  2000, \mn@doi [\icarus] {10.1006/icar.2000.6493}, \href
  {https://ui.adsabs.harvard.edu/abs/2000Icar..148...21S} {148, 21}

\bibitem[\protect\citeauthoryear{{Tamuz}, {Mazeh}  \& {Zucker}}{{Tamuz}
  et~al.}{2005}]{tamuz2005correcting}
{Tamuz} O.,  {Mazeh} T.,   {Zucker} S.,  2005, \mn@doi [\mnras]
  {10.1111/j.1365-2966.2004.08585.x}, \href
  {https://ui.adsabs.harvard.edu/abs/2005MNRAS.356.1466T} {356, 1466}

\bibitem[\protect\citeauthoryear{{Tonry} et~al.,}{{Tonry} et~al.}{2018}]{ATLAS}
{Tonry} J.~L.,  et~al., 2018, \mn@doi [\pasp] {10.1088/1538-3873/aabadf}, \href
  {https://ui.adsabs.harvard.edu/abs/2018PASP..130f4505T} {130, 064505}

\bibitem[\protect\citeauthoryear{{Warner}}{{Warner}}{2010}]{Warner2010}
{Warner} B.~D.,  2010, Minor Planet Bulletin, \href
  {https://ui.adsabs.harvard.edu/abs/2010MPBu...37..112W} {37, 112}

\bibitem[\protect\citeauthoryear{Warner, Harris  \& Pravec}{Warner
  et~al.}{2009}]{WARNER2009134}
Warner B.~D.,  Harris A.~W.,   Pravec P.,  2009, \mn@doi [Icarus]
  {https://doi.org/10.1016/j.icarus.2009.02.003}, 202, 134

\bibitem[\protect\citeauthoryear{{Waszczak} et~al.,}{{Waszczak}
  et~al.}{2015}]{2015AJ....150...75W}
{Waszczak} A.,  et~al., 2015, \mn@doi [\aj] {10.1088/0004-6256/150/3/75}, \href
  {https://ui.adsabs.harvard.edu/abs/2015AJ....150...75W} {150, 75}

\bibitem[\protect\citeauthoryear{{Xin} et~al.,}{{Xin} et~al.}{2020}]{Xin2020}
{Xin} Y.-X.,  et~al., 2020, \mn@doi [Research in Astronomy and Astrophysics]
  {10.1088/1674-4527/20/9/149}, \href
  {https://ui.adsabs.harvard.edu/abs/2020RAA....20..149X} {20, 149}

\bibitem[\protect\citeauthoryear{Zhan}{Zhan}{2021}]{zhan2021}
Zhan H.,  2021, \mn@doi [Chinese Science Bulletin]
  {https://doi.org/10.1360/TB-2021-0016}, 66, 1290

\bibitem[\protect\citeauthoryear{{{\v{D}}urech}, {Hanu{\v{s}}}, {Oszkiewicz}
  \& {Van{\v{c}}o}}{{{\v{D}}urech} et~al.}{2016}]{Durech2016}
{{\v{D}}urech} J.,  {Hanu{\v{s}}} J.,  {Oszkiewicz} D.,   {Van{\v{c}}o} R.,
  2016, \mn@doi [\aap] {10.1051/0004-6361/201527573}, \href
  {https://ui.adsabs.harvard.edu/abs/2016A&A...587A..48D} {587, A48}

\makeatother
\end{thebibliography}

% Don't change these lines
\bsp	% typesetting comment
\label{lastpage}
\end{document}